\documentclass[pra,superscriptaddress,twocolumn,english, longbibliography, showpacs=false, nofootinbib]{revtex4-2}

\usepackage[T1]{fontenc}
\usepackage[utf8]{inputenc}
\usepackage{xcolor}
\usepackage{babel}
\usepackage{amsmath}
\usepackage{amssymb}
\usepackage{subfigure}
\usepackage{graphicx}
\usepackage{physics} 
\usepackage{dsfont}
\usepackage{hyperref}
\usepackage[colorinlistoftodos, color=green!40, prependcaption]{todonotes}
\usepackage{amsthm}
\usepackage{mathtools}
\usepackage[left=23mm,right=13mm,top=35mm,columnsep=15pt]{geometry} 
\usepackage{adjustbox}
\usepackage{placeins}
\usepackage{csquotes}
\usepackage{mathrsfs}

\begin{document}

%%%%%%%%%%%%%%%%%%%%%%%%%%%%%%%%%%%%%%%%%%%%%%%%%%%%
%%%%%%%%%%%%%%%%%%%%%%%%%%%%%%%%%%%%%%%%%%%%%%%%%%%%
\title{Decoherence of a composite particle induced by a weak quantized gravitational field}

\author{Thiago H. Moreira}
\email{thiagohenriquemoreira@discente.ufg.br}
\affiliation{QPequi Group, Institute of Physics, Federal University of Goi\'as, Goi\^ania, Goi\'as, 74.690-900, Brazil}

\author{Lucas C. Céleri}
\email{lucas@qpequi.com}
\affiliation{QPequi Group, Institute of Physics, Federal University of Goi\'as, Goi\^ania, Goi\'as, 74.690-900, Brazil}

\begin{abstract}
Despite the fact that we have some proposals for the quantum theory of gravity like string theory or loop quantum gravity, we do not have any experimental evidence supporting any of these theories. Actually, we do not have experimental evidence pointing in the direction that we really need a quantum description of the gravitational field. In this scenario, several proposals for experimentally investigating quantum gravitational effects far from Plank scale have recently appear in literature, like gravitationally induced entanglement, for instance. An important issue of theses approaches is the decoherence introduced by the quantum nature not only of the system under consideration, but also from the gravitational field itself. Here, by means of the Feynman-Vernon influence functional we study the decoherence of a quantum system induced by the quantized gravitational field and by its own quantum nature. Our results may be important in providing a better understanding of many phenomena like the decoherence induced by the gravitational time-dilation, the quantum reference frames and the quantum equivalence principle.  
\end{abstract}

\maketitle
%%%%%%%%%%%%%%%%%%%%%%%%%%%%%%%%%%%%%%%%%%%%%%%%%%%%
%%%%%%%%%%%%%%%%%%%%%%%%%%%%%%%%%%%%%%%%%%%%%%%%%%%%

%%%%%%%%%%%%%%%%%%%%%%%%%%%%%%%%%%%%%%%%%%%%%%%%%%%%%%
%%%%%%%%%%%%%%%%%%%%%%%%%%%%%%%%%%%%%%%%%%%%%%%%%%%%%%
%%%%%%%%%%%%%%%%%%%%%%%%%%%%%%%%%%%%%%%%%%%%%%%%%%%%%%
%%%%%%%%%%%%%%%%%%%%%%%%%%%%%%%%%%%%%%%%%%%%%%%%%%%%%%
\section{Introduction}

Ever since the development of quantum mechanics and general relativity in the first decades of the last century, the construction of a successful quantum theory of gravity poses a problem to this day. Moreover, we do not have experimental results to guide such a construction since quantum gravitational effects are expected to be relevant at the Planck scale, which is far beyond our current technology. However, in recent years there has been increasing interest in attempts to probe the quantum nature of gravity in nonrelativistic quantum systems. The idea is not to give a full description of quantum gravitational effects, but to establish the necessity or not to quantize the field and, at the same time, to provide some experimental guidance for the theoretical construction of the theory. Among these developments we can mention investigations regarding the possibility of gravitationally mediated entanglement~\cite{Bose2017,Marletto2017,Danielson_2022,Christodoulou2023,Christodoulou2023b}, the consequences of the supperposition of spacetimes~\cite{Giacomini2022} and the study of indefinite causal order in the gravitational scenario~\cite{Moller2021,Moller2023}, just to mention a few. We point the reader to Ref.~\cite{Huggett2022} for more references. Recently another proposal has been put forward in Ref.~\cite{Parikh2020,Parikh_2021,Parikh2021} with the goal to investigate the quantum nature of the gravitational field by detecting the effect of the quantum noise induced by the gravitons on classical particles in the same spirit of the quantum Brownian motion. 

Gravitons appear as a consequence of the quantization of gravitational waves in a weak-field, perturbative regime. Although far from quantum gravity at the Planck scale, the detection of its effects on matter would provide enough evidence to establishing the true necessity of the quantization of gravity, even without a complete theory. After the observations of gravitational waves announced by LIGO~\cite{Abbott_2016}, questions have been raised on the possibility that these could also reveal quantum effects of the gravitational field. However, according to arguments by F. Dyson~\cite{Dyson_2013}, the detection of individual gravitons was bound to be nearly impossible. 

The proposal in Refs.~\cite{Parikh2020,Parikh_2021,Parikh2021} focus instead on detecting the effect on falling bodies due to the quantization of the gravitational field, which enters the equation of motion as a kind of quantum noise. Employing the Feynman-Vernon influence functional formalism~\cite{Feynman1963,Feynman2010}, the authors obtained a Langevin-like stochastic equation describing the geodesic deviation of two falling masses with the gravitons effect entering as a stochastic force. The noise kernel was computed for four distinct classes of initial graviton states, namely the Minkowski vacuum as well as thermal, coherent and squeezed states, with the latter exhibiting an exponential enhancement of the effect. The same analysis was performed in Ref.~\cite{Kanno2021} by means of a different approach and later on it was also done in Ref.~\cite{Cho2022}, where all graviton modes and polarizations were taken into account.

On the other hand, there has also been an increasing interest on investigating decoherence induced by gravitational fields, both classical and quantum (for a review, see Ref.~\cite{Bassi_2017} and references therein). By gravitational decoherence we mean a setup in which gravitational perturbations are viewed as an environment affecting the evolution of quantum particles in superposition in a given basis that leads to the loss of quantum coherence~\cite{Anastopoulos_2013,Vedral-2020-Arxiv,Blencowe_2013,Suzuki_2015}. There have also been investigations on the decoherence of a composite particle~\cite{Zych2019} induced by the entanglement of the center-of-mass variables with internal degrees of freedom caused by gravitational time dilation~\cite{Pikovski2015,Pikovski2017} and, more recently, in any given spacetime manifold described by a generic Riemann curvature tensor~\cite{Singh-2023-Arxiv}.

In this scenario, we study the decoherence of a particle that contains both external and internal degrees of freedom that are coupled by the presence of a quantized weak gravitational field. We focus on the decoherence of the external degrees of freedom, that will be affected by the internal ones and by the quantum fluctuations of the gravitational field. We treat the problem by using the Feynman-Vernon influence functional formalism of open quantum systems, which is closely related to the closed-time-path integral method~\cite{Feynman1963,Feynman2010,Johnson_2002,Breuer2001,Hu_1992,Bonan_a_2006,Calzetta2008}. We follow the analysis of decoherence induced by gravitons discussed in Ref.~\cite{Kanno2021} with the difference that we are considering a particle whose internal degrees of freedom are described quantum mechanically. Since the gravitational field couples with all degrees of freedom in the same way, we are effectively dealing with an open quantum system interacting with two interacting environments. This becomes important in the quest for a deeper understanding of several phenomena involving quantum systems and the gravitational field, among which we can mention the decoherence induced by the gravitational time-dilation~\cite{Pikovski2015,Pikovski2017} and the quantum equivalence principle~\cite{Zych2018}.

The article is organized as follows. In Sec.~\ref{S:Gravitational-influence-functional} we discuss the total action of the system and, by treating the field as a quantum environment we obtain the reduced density matrix of the relevant variables in terms of the gravitational influence functional. Since the total action is quadratic, we can solve the problem exactly by encoding all the gravitational influence into the noise and dissipation kernels. The internal structure of the system is taken into account in Sec.~\ref{S:The-external-degrees-of-freedom-density-matrix}, where we split the particle degrees of freedom into external and internal variables, and integrate the internal ones by applying the influence functional method once more. In order to make this calculation possible, we are considering such degrees of freedom to be described by independent harmonic oscillators in thermal equilibrium. This leads to an expression for the reduced density matrix of the external degrees of freedom of the quantum particle. In Sec.~\ref{S:Decoherence-rate} we take the external degrees of freedom to be in a superposition of two distinct paths and compute the decoherence rate, as well as the decoherence time, by considering different initial states for the quantum gravitational field. Our conclusions and some final remarks are left to Sec.~\ref{S:Conclusions}. Details of the calculations are presented in the appendices. 

Unless stated otherwise, we work in Planck units by setting $\hbar=c=G=k_B=1$. The metric signature is $(-+++)$. Greek indexes in tensors and vectors runs from $0$ to $3$, while Latin ones takes the spacial labels $1$, $2$ and $3$. 

%%%%%%%%%%%%%%%%%%%%%%%%%%%%%%%%%%%%%%%%%%%%%%%%%%
%%%%%%%%%%%%%%%%%%%%%%%%%%%%%%%%%%%%%%%%%%%%%%%%%%
%%%%%%%%%%%%%%%%%%%%%%%%%%%%%%%%%%%%%%%%%%%%%%%%%%
%%%%%%%%%%%%%%%%%%%%%%%%%%%%%%%%%%%%%%%%%%%%%%%%%%
\section{Gravitational influence functional} \label{S:Gravitational-influence-functional}

The goal of this section is twofold. First, we briefly review the results obtained in Refs.~\cite{Parikh2020,Parikh_2021,Parikh2021}, presenting the system we are considering as well as setting the notation that will be employed in the rest of the paper. This will make the present paper self-contained. Additionally, we add to this formalism the quantum description of the internal degrees of freedom of our system of interest. Specifically, we consider two freely falling particles on a spacetime manifold over which a metric $g_{\mu\nu}$ is defined. Our aim here is to establish the response of the particles to the quantized gravitational radiation field in order to set up the basis for the decoherence analysis we present in the sequence.

%%%%%%%%%%%%%%%%%%%%%%%%%%%%%%%%%%%%%%%%%%%%%%%%%%%%%%
%%%%%%%%%%%%%%%%%%%%%%%%%%%%%%%%%%%%%%%%%%%%%%%%%%%%%%
\subsection{The classical action}

Let us start by writing the classical action of a weak gravitational field coupled to a pair of free falling massive particles. We write the total action as $S = S_{\rm matter} + S_{\rm EH}$, with the first term containing the action of the particles and their interaction with the gravitational field, while the second term describes the Einstein-Hilbert action of the field.

If the coordinates of the particles are denoted by $\zeta^\mu$ and $\xi^\mu$, the action for both particles is given by
\begin{eqnarray}\label{S_detector-general}
    S_{\rm matter}&=& -M_{0}\int \dd t\,\sqrt{-g_{\mu\nu}\dot{\zeta}^\mu\dot{\zeta}^\nu} \nonumber\\
    &+&\int \dd t\,L_{\rm rest}\sqrt{-g_{\mu\nu}\dot{\xi}^\mu\dot{\xi}^\nu},
\end{eqnarray}
where a dot means differentiation with respect to the coordinate time $t$. Since the first particle has no internal degrees of freedom we simply choose its rest Lagrangian as its mass $M_0$. The second particle is considered to be structured, with external and internal degrees of freedom, in such a way we can write its rest Lagrangian as
    \begin{equation} \label{Rest-Lagrangian}
        L_{\rm rest}(\varrho,\dot{\varrho}\,\bar{t})=-m_0+\lambda L_{\rm int}(\varrho,\dot{\varrho}\,\bar{t}).
    \end{equation}
In this equation, $m_{0}$ is the mass of the second particle, while $L_{\rm int}(\varrho,\dot{\varrho}\,\bar{t})$ describes its internal degrees of freedom with coordinates $\varrho$ and generalised velocities $\dot{\varrho}=\dv*{\varrho}{t}$. Also, $\bar{t}=\dv*{t}{\tau}$ with $\tau$ being the particle's proper time and $\lambda$ is a dimensionless parameter introduced for mathematical convenience. Additionally, we consider that the first particle is on shell with worldline $\zeta_0^\mu(t)$ and $M_0 \gg L_{\rm rest}$. Furthermore, we place the first particle at rest at the origin of our coordinate system, $\zeta_{0}^\mu(t)=t\,\delta_0^\mu$, such that the coordinate time $t$ is the first particle's proper time. Since we will focus on the second particle, from here on we simply refer to it as the system.

Under such assumptions, the first term in the action~\eqref{S_detector-general} essentially has no dynamics. It is then appropriate to think of $(t,\xi^i)$ as the Fermi normal coordinates defined with respect to the worldline of the first particle. In these coordinates, we can write the metric components as~\cite{Manasse1963}
\begin{equation}
    \begin{split}
        g_{00}(t,\xi^i)&=-1-R_{i0j0}(t,0)\xi^i\xi^j+O(\xi^3), \\
        g_{0i}(t,\xi^i)&=-\frac{2}{3}R_{0jik}(t,0)\xi^j\xi^k+O(\xi^3), \\
        g_{ij}(t,\xi^i)&=\delta_{ij}-\frac{1}{3}R_{ikjl}(t,0)\xi^k\xi^l+O(\xi^3),
    \end{split}
\end{equation}
where $R_{\mu\nu\rho\sigma}$ is the Riemann tensor. In our parameterization, $\xi^0(t)=t$ and thus
\begin{align} \label{dot-tau}
    \dot{\tau}&\equiv\sqrt{-g_{\mu\nu}(\xi)\dot{\xi}^\mu\dot{\xi}^\nu}\simeq\sqrt{1+R_{i0j0}(t,0)\xi^i\xi^j-\delta_{ij}\dot{\xi}^i\dot{\xi}^j} \nonumber \\
    &\simeq1-\frac{1}{2}\delta_{ij}\dot{\xi}^i\dot{\xi}^j+\frac{1}{2}R_{i0j0}(t,0)\xi^i\xi^j,
\end{align}
in the nonrelativistic limit. 

As usual, we express the weakness of the gravitational field by decomposing the metric into the flat Minkowski metric $\eta_{\mu\nu}=\rm{diag}(-1,1,1,1)$ and a small perturbation $g_{\mu\nu}=\eta_{\mu\nu}+h_{\mu\nu}$, with $\abs{h_{\mu\nu}}\ll 1$~\nocite{Carroll}.

Now, by taking advantage of the gauge invariance of linearized gravity, we choose to work in the transverse-traceless gauge such that the metric perturbation obeys $\Bar{h}_{0\nu}=0$, $\eta^{\mu\nu}\Bar{h}_{\mu\nu}=0$ and $\partial^\mu\Bar{h}_{\mu\nu}=0$, with $\Bar{h}_{\mu\nu} = h_{\mu\nu} - h\,\eta_{\mu\nu}/2$ and $h = h^{\mu}_{\mu}$~\cite{Carroll}. In this gauge, the non-vanishing components of the Riemann curvature tensor are given by $R_{0i0j}(t,0)=-\ddot{\Bar{h}}_{ij}(t,0)/2$, thus implying that Eq.~\eqref{dot-tau} takes the form
\begin{equation*}
    \dot{\tau}\simeq1-\frac{1}{2}\qty[\delta_{ij}\dot{\xi}^i\dot{\xi}^j+\frac{1}{2}\ddot{\Bar{h}}_{ij}(t,0)\xi^i\xi^j],
\end{equation*}
which, by dropping the non-dynamical terms, results in the matter action
\begin{eqnarray}\label{Matter-action}
    S_{\rm matter}  &=&\lambda\int \dd t\,L_{\rm int}(\varrho,\dot{\varrho}\bar{t}) \nonumber\\
    &+& \frac{1}{2}\int \dd t\,\qty[m_0-\lambda L_{\rm int}(\varrho,\dot{\varrho}\bar{t})]\,\delta_{ij}\dot{\xi}^i\dot{\xi}^j \nonumber \\
    &+&\frac{1}{4}\int \dd t\,\qty[m_0-\lambda L_{\rm int}(\varrho,\dot{\varrho}\bar{t})]\,\ddot{\Bar{h}}_{ij}(t,0)\xi^i\xi^j.
\end{eqnarray}
Note that the first two terms of this equation represents the free action (including the interaction between the internal and external degrees of freedom) of the system in the Minkowisk background, while the last one represents the interaction of the system with the gravitational perturbation.

We now move to the gravitational field, which is described by the Einstein-Hilbert action~\cite{Carroll}
\begin{equation*}
    S_{\rm EH}=\frac{1}{16\pi}\int \dd^4x\,\sqrt{-g}\,R,
\end{equation*}
where $g$ stands for the determinant of the metric and $R$ is the Ricci scalar. Now, up to second order in the perturbation and in the transverse-traceless gauge, this action reduces to
\begin{equation}\label{Eistein-Hilbert-action-TT-gauge}
    S_{\rm EH} = -\frac{1}{64\pi}\int \dd^4x\,\partial_\mu\Bar{h}_{ij}\partial^\mu\Bar{h}^{ij},
\end{equation}
from which we immediately obtain the wave equation $\Box\Bar{h}_{ij}=0$. We may then express the gravitational waves in terms of its two physical degrees of freedom corresponding to the two polarization components by means of a Fourier transform~\cite{Carroll,Cho2022}, 
\begin{equation} \label{Fourier-transform}
    \Bar{h}_{ij}(t,\vb{x})=\int \dd^3k\,\sum_s\epsilon_{ij}^{s}(\vb{k})q_s(t,\vb{k})e^{i\vb{k}\cdot\vb{x}},
\end{equation}
where $\epsilon_{ij}^{s}$ denotes the polarization tensor, which satisfies the normalization $\epsilon_{ij}^s(\vb{k})\epsilon^{ij}_{s'}(\vb{k})=2\delta^s_{s'}$, transversality $k^i\epsilon_{ij}^{s}(\vb{k})=0$ and traceless $\delta^{ij}\epsilon_{ij}^{s}(\vb{k})=0$ conditions. By plugging Eq.~\eqref{Fourier-transform} into Eq.~\eqref{Eistein-Hilbert-action-TT-gauge} we obtain
\begin{equation*}
    S_{\rm EH}=\frac{\pi^{2}}{8}\int dt\int \dd^3k\,\sum_s\qty[\abs{\dot{q}_s(t,\vb{k})}^2-\vb{k}^2\abs{q_s(t,\vb{k})}^2],
\end{equation*}
where we used the fact that $\Bar{h}_{ij}$ must be real, thus implying that $\epsilon_{ij}^{s}(\vb{k})q_s^*(t,\vb{k})=\epsilon_{ij}^{s}(-\vb{k})q_s(t,-\vb{k})$.

Before writing down the total action, we note that
\begin{align*}
    \ddot{\Bar{h}}_{ij}(t,0)\xi^i\xi^j&=\int \dd^3k\,\sum_s\epsilon_{ij}^s(\vb{k})\ddot{q}_s(t,\vb{k})\xi^i\xi^j \\
    &=\frac{1}{2}\int \dd^3k\,\sum_s\epsilon_{ij}^s(\vb{k})\qty[\ddot{q}_s(t,\vb{k})+\ddot{q}^*_s(t,\vb{k})]\xi^i\xi^j \\
    &=\int \dd^3k\,\sum_s\epsilon_{ij}^s(\vb{k})\textrm{Re}[\ddot{q}_s(t,\vb{k})]\xi^i\xi^j,
\end{align*}
which follows from Eqs.~\eqref{Matter-action} and~\eqref{Fourier-transform}. We then see that only the real part of the amplitude couples to the matter degrees of freedom. We therefore take these amplitudes to be real and write the total action as
\begin{align}
    S&=S_{\rm EH}+S_{\rm p}+S_{\rm int}\nonumber \\
    &=\frac{\pi^{2}}{8}\int \dd t\int \dd^3k\,\sum_s\qty[\dot{q}^2_s(t,\vb{k})-\vb{k}^2q_s^2(t,\vb{k})] \nonumber \\
    &\hspace{0.5cm}+\frac{1}{2}\int \dd t\,\qty[m_0-\lambda L_{\rm int}(\varrho,\dot{\varrho}\bar{t})]\,\delta_{ij}\dot{\xi}^i\dot{\xi}^j \nonumber \\
    &\hspace{0.5cm}+\lambda\int \dd t\,L_{\rm int}(\varrho,\dot{\varrho}\bar{t}) \nonumber \\
    &\hspace{0.5cm}+\frac{m_0}{4}\int \dd t\,\int \dd^3k\,\sum_sq_s(t,\vb{k})\,X^s(t,\vb{k}),
\end{align}
where we have defined
\begin{equation} \label{X(t)-definition}
    X^s(t,\vb{k})\equiv\dv[2]{t}\qty{\epsilon_{ij}^s(\vb{k})\xi^i\xi^j\qty[1-\frac{\lambda}{m_0}L_{\rm int}(\varrho,\dot{\varrho}\bar{t})]}.
\end{equation}

Our goal now is to quantize the gravitational field and to study the effect of its quantum fluctuations on the system of interest.

%%%%%%%%%%%%%%%%%%%%%%%%%%%%%%%%%%%%%%%%%%%%%%%%%%%%%%%%%%%
%%%%%%%%%%%%%%%%%%%%%%%%%%%%%%%%%%%%%%%%%%%%%%%%%%%%%%%%%%%
\subsection{Stochastic effective action}

The interaction between the system and the gravitational field gives rise to a coupling between the system's external and internal variables. Ultimately, we are not interested in what happens to the gravitons themselves, and the quantized gravitational field shall be treated as an environment. We treat this problem by using the Feynman-Vernon influence functional approach to open quantum systems~\cite{Feynman1963,Feynman2010}. In this approach, the influence functional is defined as
\begin{equation}
    \mathcal{F}[x,x']=e^{iS_{\rm IF}[x,x',t]},
\end{equation}
where $x$ and $x'$ are distinct values of the system variables and $S_{\rm IF}$ is the influence action.

Taking into account that the gravitational part of our system is quadratic in the relevant variables and the interaction is linear, we obtain the following result
\begin{widetext}
\begin{equation} \label{S_IF}
\begin{split}
    S_{\rm IF}^{g}[x,x']=\int \dd t\dd t\left\{ \frac{1}{2}\qty[x_{ij}(t)-x'_{ij}(t)]D^{ijkl}_{\rm g}(t,t')\qty[x_{kl}(t')+x'_{kl}(t')] \right. \\
    \left. +\frac{i}{2}\qty[x_{ij}(t)-x'_{ij}(t)]N^{ijkl}_{\rm g}(t,t')\qty[x_{kl}(t')-x'_{kl}(t')]\right\} ,
\end{split}
\end{equation}
where we have defined
\begin{equation} \label{x-ij(t)-definition}
    x_{ij}(t)\equiv\xi_i(t)\xi_j(t)\qty[1-\frac{\lambda}{m_0}L_{\rm int}(\varrho,\dot{\varrho}\bar{t})],
\end{equation}
while
\begin{subequations}
    \begin{equation}
        D^{ijkl}_{\rm g}(t,t')=i\qty(\frac{m_{0}}{4})^2\dv[2]{t}\dv[2]{{t'}}\int \dd^3k\,\sum_s\epsilon^{ij}_s(\vb{k})\epsilon^{kl}_s(\vb{k})\expval{\comm{q_s(t,\vb{k})}{q_s(t',\vb{k})}}_{\rm g}\theta(t-t'),
    \end{equation}
    and
    \begin{equation} \label{Noise-kernel-definition}
        N^{ijkl}_{\rm g}(t,t')=\frac{1}{2}\qty(\frac{m_{0}}{4})^2\dv[2]{t}\dv[2]{{t'}}\int \dd^3k\,\sum_s\epsilon^{ij}_s(\vb{k})\epsilon^{kl}_s(\vb{k})\expval{\acomm{q_s(t,\vb{k})}{q_s(t',\vb{k})}}_{\rm g},
    \end{equation}
\end{subequations}
\end{widetext}
are the (gravitational) dissipation and noise kernels, respectively~\cite{Cho2022,Calzetta2008}. The $q$'s stand for position operators in the Heisenberg representation. The expectation values with the subscript $g$ are computed with respect to the initial state of the gravitons. Also, in deriving Eq.~\eqref{S_IF} we have used the fact that each graviton mode (and polarization) is independent of all the others and therefore can be treated separately in such a way that the total influence action is the sum of the action corresponding to each mode (and polarization)~\cite{Feynman1963,Feynman2010}. The details of the calculations are presented in Appendix~\ref{A:Quantum-open-systems-and-the-Feynman-Vernon-influence-functional}.

Following Feynman and Vernon, we express the noise term in $S_{\rm IF}^{g}$ in terms of a stochastic variable $\mathcal{N}_{ij}(t)$ using the Gaussian functional identity~\cite{Cho2022}
\begin{equation}
\begin{split}
    &e^{-\frac{1}{2}\int dtdt'\,\qty[x_{ij}(t)-x'_{ij}(t)]N^{ijkl}_{\rm g}(t,t')\qty[x_{kl}(t')-x'_{kl}(t')]} \\
    &\hspace{0.2cm}=\mathcal{C}\int\mathcal{D}\mathcal{N}\,e^{-\frac{1}{2}\int dtdt'\,\mathcal{N}_{ij}(t)(N_g^{-1})^{ijkl}(t,t')\mathcal{N}_{kl}(t')} \\
    &\hspace{1.5cm}\times e^{-i\int dt\,\mathcal{N}^{ij}(t)\qty[x_{ij}(t)-x'_{ij}(t)]},
\end{split}
\end{equation}
where $\mathcal{C}$ is a normalization constant. Stochastic averages are performed considering the Gaussian probability density
\begin{equation}
    P[\mathcal{N}]=\mathcal{C}\,e^{-\frac{1}{2}\int dtdt'\,\mathcal{N}_{ij}(t)(N_g^{-1})^{ijkl}(t,t')\mathcal{N}_{kl}(t')}.
\end{equation}
For example, the one- and two-point correlation functions are given by
\begin{subequations}  \label{Stochastic-averages}
\begin{equation}
    \expval{\mathcal{N}^{ij}(t)}_{\rm sto}=\int\mathcal{D}\mathcal{N}\,P[\mathcal{N}]\mathcal{N}^{ij}(t)=0,
\end{equation}
\begin{align}
    \expval{\mathcal{N}^{ij}(t)\mathcal{N}^{kl}(t')}_{\rm sto}&=\int\mathcal{D}\mathcal{N}\,P[\mathcal{N}]\mathcal{N}^{ij}(t)\mathcal{N}^{kl}(t') \nonumber \\
    &=N_{\rm g}^{ijkl}(t,t').
\end{align}
\end{subequations}

The (gravitational) influence functional then becomes
\begin{equation} \label{Influence-functional-final}
\begin{split}
    &e^{iS_{\rm IF}^{g}[x,x']} \\
    &\hspace{0.1cm}=\int\mathcal{D}\mathcal{N}\,P[\mathcal{N}]\,e^{-i\int dt\,\mathcal{N}^{ij}(t)\qty[x_{ij}(t)-x'_{ij}(t)]} \\
    &\hspace{0.5cm}\times e^{\frac{i}{2}\int dtdt'\qty[x_{ij}(t)-x'_{ij}(t)]D^{ijkl}_{\rm g}(t,t')\qty[x_{kl}(t')+x'_{kl}(t')]}.
\end{split}
\end{equation}

We now proceed by considering only the external degrees of freedom of our system since we are interested in the action of the noise coming from its internal degrees of freedom and from the quantum fluctuations of the gravitational field. In order to do this, we have to compute the reduced density matrix of the relevant degrees of freedom while tracing out all the others.

%%%%%%%%%%%%%%%%%%%%%%%%%%%%%%%%%%%%%%%%%%%%%%%%%%%%%%%%%%%%%%
%%%%%%%%%%%%%%%%%%%%%%%%%%%%%%%%%%%%%%%%%%%%%%%%%%%%%%%%%%%%%%
%%%%%%%%%%%%%%%%%%%%%%%%%%%%%%%%%%%%%%%%%%%%%%%%%%%%%%%%%%%%%%
\section{The external degrees of freedom density matrix} \label{S:The-external-degrees-of-freedom-density-matrix}

The particle (external + internal degrees of freedom) density matrix at time $t$ is given by [see Eq.~\eqref{Evolution-of-reduced-density-matrix}]
\begin{equation} \label{Total-density-matrix-of-system}
\begin{split}
    \rho_{\rm p}(\xi,\varrho,\xi',\varrho,t)=\int \dd\xi(0)\dd\xi'(0)\dd\varrho(0)\dd\varrho'(0) \\
    \times\mathcal{J}(\xi,\varrho,\xi',\varrho,t|\xi(0),\varrho(0),\xi'(0),\varrho'(0),0) \\
    \times\rho_{\rm p}(\xi(0),\varrho(0),\xi'(0),\varrho'(0),0),
\end{split}
\end{equation}
where
\begin{equation} \label{Evolution-J}
\begin{split}
    &\mathcal{J}(\xi,\varrho,\xi',\varrho,t|\xi(0),\varrho(0),\xi'(0),\varrho'(0),0)) \\
    &=\int\mathcal{D}\xi\mathcal{D}\xi'\mathcal{D}\varrho\mathcal{D}\varrho'\,e^{i\qty(S_{\rm sys}[\xi,\varrho]-S_{\rm sys}[\xi',\varrho'])}e^{iS^{g}_{\rm IF}[\xi,\varrho,\xi',\varrho']},
\end{split}
\end{equation}
with $e^{iS^{g}_{\rm IF}}$ given in Eq.~\eqref{Influence-functional-final} in terms of the variable $x_{ij}(t)$ that is defined in Eq.~\eqref{x-ij(t)-definition}.

We observe that the term involving the dissipation kernel in Eq. \eqref{Influence-functional-final} is of order $O(\xi^4)$ since $x_{ij}$ is already of order $O(\xi^2)$. Thus the leading order contribution comes from the term involving the noise kernel and we approximate Eq.~\eqref{Influence-functional-final} as
\begin{equation}
    e^{iS_{\rm IF}^{g}[x,x']}\simeq\int\mathcal{D}\mathcal{N}\,P[\mathcal{N}]\,e^{-i\int dt\,\mathcal{N}^{ij}(t)\qty[x_{ij}(t)-x'_{ij}(t)]}.
\end{equation}
This means that we are only considering the influence of the gravitational field encoded in the noise kernel, which depends on the initial state of the gravitons, as can be seen from the definition \eqref{Noise-kernel-definition}.

Within the above approximation we find
\begin{equation}
\begin{split}
    &\mathcal{J}(\xi,\varrho,\xi',\varrho,t|\xi(0),\varrho(0),\xi'(0),\varrho'(0),0)) \\
    &=\int\mathcal{D}\xi\mathcal{D}\xi'\mathcal{D}\varrho\mathcal{D}\varrho'\mathcal{D}\mathcal{N}\,P[\mathcal{N}]\,e^{i\qty(S_{\rm eff}[\xi,\varrho,\mathcal{N}]-S_{\rm eff}[\xi',\varrho',\mathcal{N}])},
\end{split}
\end{equation}
where
\begin{equation}
\begin{split}
    &S_{\rm eff}[\xi,\varrho,\mathcal{N}]=\int dt\,\left[ \frac{1}{2}m_0\delta_{ij}\dot{\xi}^i\dot{\xi}^j-\mathcal{N}_{ij}\xi^i\xi^j\right. \\
    &\hspace{0.2cm}\left.-\lambda\qty(\frac{1}{2}\delta_{ij}\dot{\xi}^i\dot{\xi}^j-\frac{1}{m_0}\mathcal{N}_{ij}\xi^i\xi^j-1)L_{\rm int}(\varrho,\dot{\varrho}\bar{t})\right] .
\end{split}
\end{equation}

We now assume that initially the external and internal degrees of freedom of our system are uncorrelated, thus implying that 
\begin{equation*}
    \rho_{\rm p}\left(\xi,\varrho,\xi',\varrho',0\right)
 =\rho_{\rm ext}\left(\xi,\xi',0)\rho_{\rm int}(\varrho,\varrho',0\right),
\end{equation*}
where $\rho_{\rm ext}$ ($\rho_{\rm int}$) stands for the external (internal) degrees of freedom density matrix elements and all the variables are written at the initial time. The interaction with the gravitons couples the external and internal variables and we are left with the total density matrix \eqref{Total-density-matrix-of-system}. We are interested here in the dynamics of the external degrees of freedom and, thus, we proceed by taking the partial trace over the internal ones
\begin{equation*}
    \rho_{\rm ext}(\xi,\xi',t)=\int \dd\varrho\,\rho_{\rm p}(\xi,\varrho,\xi',\varrho,t).
\end{equation*}
resulting in
\begin{equation}
\begin{split}
    \rho_{\rm ext}&(\xi,\xi',t)=\int \dd\xi(0)d\xi'(0)\,\rho_{\rm ext}(\xi(0),\xi'(0),0) \\
    &\times\int\mathcal{D}\mathcal{N}\,P[\mathcal{N}]\int_{\xi(0)}^\xi\mathcal{D}\xi\int_{\xi'(0)}^{\xi'}\mathcal{D}\xi' \\
    &\times e^{i\qty(S_{\rm eff}^{(1)}[\xi,\mathcal{N}]-S_{\rm eff}^{(1)}[\xi',\mathcal{N}])}e^{S_{\rm IF}^{(\textrm{int})}[\xi,\xi',\mathcal{N}]},
\end{split}
\end{equation}
where
\begin{equation} \label{Internal-DoF-IF}
\begin{split}
    &e^{iS_{\rm IF}^{(\textrm{int})}[\xi,\xi',\mathcal{N}]} \\
    \hspace{0.2cm}&=\int d\varrho d\varrho(0)d\varrho'(0)\,\rho_{\rm int}(\varrho(0),\varrho'(0),0) \\
    &\hspace{0.2cm}\times\int_{\varrho(0)}^\varrho\mathcal{D}\varrho\int_{\varrho'(0)}^\varrho\mathcal{D}\varrho'\,e^{i\qty(S_{\rm eff}^{(2)}[\xi,\varrho,\mathcal{N}]-S_{\rm eff}^{(2)}[\xi',\varrho',\mathcal{N}])},
\end{split}
\end{equation}
while
\begin{subequations}
\begin{equation}
    S_{\rm eff}^{(1)}=\int dt\qty(\frac{1}{2}m_0\delta_{ij}\dot{\xi}^i\dot{\xi}^j-\mathcal{N}_{ij}\xi^i\xi^j),
\end{equation}
and
\begin{equation} \label{S-eff-(2)}
    S_{\rm eff}^{(2)}=\lambda\int dt\qty(1-\frac{1}{2}\delta_{ij}\dot{\xi}^i\dot{\xi}^j+\frac{1}{m_0}\mathcal{N}_{ij}\xi^i\xi^j)L_{\rm int}(\varrho,\dot{\varrho}\bar{t}).
\end{equation}
\end{subequations}

Thus, we essentially have the same problem as the one treated in Sec.~\ref{S:Gravitational-influence-functional}, namely a system interacting with a quantum environment. Therefore, we shall compute the Feynman-Vernon influence functional once again. It is worth to remark that the total influence functional (gravitons plus internal degrees of freedom) is not simply the product of the individual functionals, since the gravitational field couples with all variables describing the system. When considering the system of interest to be the external degrees of freedom, we effectively end with two environments that interact with the system and with each other. In such a case the additive property of the influence action for multiple environments does not hold~\cite{Feynman1963}.

In order to proceed with our calculations, let us assume that the Lagrangian describing the internal degrees of freedom is of the form
\begin{equation}
    L_{\rm int}(\varrho,\dot{\varrho}\,\bar{t})=\sum_\alpha\qty[\frac{1}{2}\mu_\alpha\Bar{\varrho}_\alpha^2-V(\varrho_\alpha)],
\end{equation}
where the $\mu$'s are some reduced masses of the system and $V$ is a function of the coordinates. Since we want to keep terms only up to second order in the position and velocity coordinates, we may write
\begin{equation}
    L_{\rm int}(\varrho,\dot{\varrho}\,\bar{t})\simeq\sum_\alpha\qty(\frac{1}{2}\mu_\alpha\dot{\varrho}_\alpha^2-\vartheta_\alpha\varrho_\alpha-\frac{1}{2}\mu_\alpha\varpi_\alpha^2\varrho_\alpha^2),
\end{equation}
where the $\vartheta$'s and the $\varpi$'s are constants. Then, Eq.~\eqref{S-eff-(2)} becomes
\begin{subequations}
\begin{equation}
\begin{split}
    S_{\rm eff}^{(2)}=\sum_\alpha\left[ \lambda\int dt\qty(\frac{1}{2}\mu_\alpha\dot{\varrho}_\alpha^2-\frac{1}{2}\mu_\alpha\varpi_\alpha^2\varrho_\alpha^2)\right. \\
    \left. +\lambda\vartheta_\alpha\int dt\,Y(t)\varrho_\alpha(t)\right] ,
\end{split}
\end{equation}
with
\begin{equation}
    Y(t)=\frac{1}{2}\delta_{ij}\dot{\xi}^i\dot{\xi}^j-\frac{1}{m_0}\mathcal{N}_{ij}\xi^i\xi^j-1.
\end{equation}
\end{subequations}

Therefore, we are considering the internal degrees of freedom to be approximated by a set of independent harmonic oscillators that couples linearly with the external degrees of freedom. In that case the internal degrees of freedom influence functional \eqref{Internal-DoF-IF} is Gaussian, implying that we can write the influence action as
\begin{equation}
\begin{split}
    &S^{(\textrm{int})}_{\rm IF}[Y,Y']=\int dtdt' \\
    &\times\left\{ \frac{1}{2}\qty[Y(t)-Y'(t)]D_{\rm int}(t,t')\qty[Y(t')+Y'(t')]\right. \\
    &\hspace{0.6cm}\left. +\frac{i}{2}\qty[Y(t)-Y'(t)]N_{\rm int}(t,t')\qty[Y(t')-Y'(t')]\right\} ,
\end{split}
\end{equation}
with
\begin{subequations}
    \begin{equation}
        D_{\rm int}(t,t')=i\lambda^2\sum_\alpha\vartheta_\alpha^2\expval{\comm{\varrho_\alpha(t)}{\varrho_\alpha(t')}}_{\rm int}\theta(t-t'),
    \end{equation}
    \begin{equation} \label{Internal-dofs-noise-kernel}
        N_{\rm int}(t,t')=\frac{1}{2}\lambda^2\sum_\alpha\vartheta_\alpha^2\expval{\acomm{\varrho_\alpha(t)}{\varrho_\alpha(t')}}_{\rm int},
    \end{equation}
\end{subequations}
where now the $\varrho(t)$'s are operators in the Heisenberg representation.

Similarly to what we did for the term containing the noise kernel for the gravitational influence functional, we can express the noise term in $S_{\rm IF}^{(\textrm{int})}$ in terms of a stochastic variable using the same Gaussian functional identity. Then, this term in the internal degrees of freedom influence functional will give place to a Gaussian probability density and a linear term on the $Y$'s variables. Therefore, just like in the gravitational case, the leading order contributions come from the noise term and we may take
\begin{equation}
\begin{split}
    &e^{iS_{\rm IF}^{(\textrm{int})}[Y,Y']} \\
    &\hspace{0.2cm}\simeq e^{-\frac{1}{2}\int dtdt'\,\qty[Y(t)-Y'(t)]N_{\rm int}(t,t')\qty[Y(t')-Y'(t')]},
\end{split}
\end{equation}
resulting in
\begin{widetext}
\begin{eqnarray} \label{rho(xi,xi',t)}
        \rho_{\rm ext}(\xi,\xi',t)&=&\int d\xi(0)d\xi'(0)\,\rho_{\rm ext}(\xi(0),\xi'(0),0) \int_{\xi(0)}^\xi\mathcal{D}\xi\int_{\xi'(0)}^{\xi'}\mathcal{D}\xi'\,e^{\frac{i}{2}m_0\delta^{ij}\int dt\,\qty(\dot{\xi}_i\dot{\xi}_j-\dot{\xi}_i'\dot{\xi}_j')}\\    &\times&\int\mathcal{D}\mathcal{N}\,P[\mathcal{N}]\,e^{-i\int dt\,\mathcal{N}^{ij}\qty(\xi_i\xi_j-\xi_i'\xi_j')}e^{-\frac{1}{2}\int dtdt'\,\qty[Y(t)-Y'(t)]N_{\rm int}(t,t')\qty[Y(t')-Y'(t')]}.
\end{eqnarray}
Fallowing the same reasoning presented in the previous section, we consider only the influence encoded into the noise kernel.

The stochastic averages shown in Eqs.~\eqref{Stochastic-averages} can be employed provided we work on a perturbative regime (dropping terms of order $\xi^3$ and higher). From this, we obtain the external degrees of freedom density matrix given as

\begin{align} \label{rho(xi,xi',t)-2}
    \rho_{\rm ext}(\xi,\xi',t)&\simeq\int d\xi(0)d\xi'(0)\,\rho_{\rm ext}(\xi(0),\xi'(0),0)\int_{\xi(0)}^\xi\mathcal{D}\xi\int_{\xi'(0)}^{\xi'}\mathcal{D}\xi'\,e^{\frac{i}{2}m_0\delta^{ij}\int dt\,\qty(\dot{\xi}_i\dot{\xi}_j-\dot{\xi}_i'\dot{\xi}_j')} \nonumber \\
    &\hspace{0.5cm}\times\exp\left\{ -\frac{1}{4}\delta^{ij}\delta^{kl}\int dtdt'\,\qty[\dot{\xi}_i(t)\dot{\xi}_j(t)-\dot{\xi}_i'(t)\dot{\xi}_j'(t)]N_{\rm int}(t,t')\qty[\dot{\xi}_k(t')\dot{\xi}_l(t')-\dot{\xi}_k'(t')\dot{\xi}_l'(t')]\right. \\
    &\hspace{0.7cm}\left. -\int dtdt'\,\qty[\xi_i(t)\xi_j(t)-\xi'_i(t)\xi'_j(t)]\qty[\frac{1}{2}+\frac{1}{m_0^2}N_{\rm int}(t,t')]N^{ijkl}_{\rm g}(t,t')\qty[\xi_k(t')\xi_l(t')-\xi'_k(t')\xi'_l(t')]\right\} . \nonumber
\end{align}
\end{widetext}

As we can see, the off-diagonal elements of the density matrix in the coordinate basis ($\xi\neq\xi'$) evolves with a exponentially decaying amplitude. Therefore, for an initial superposition state of the external degrees of freedom, the interaction with the internal degrees of freedom induced by the gravitons as well as the interaction with the gravitons themselves inevitably leads to lost of quantum coherence. In order to determine how significant these effects are, in the next section we compute the decoherence rate.

%%%%%%%%%%%%%%%%%%%%%%%%%%%%%%%%%%%%%%%%%%%%%%%%%%%%%%%%%%%%
%%%%%%%%%%%%%%%%%%%%%%%%%%%%%%%%%%%%%%%%%%%%%%%%%%%%%%%%%%%%
\section{Decoherence rate} \label{S:Decoherence-rate}

We assume that the system is in a superposition state of two spatially separated locations, $\rho_{\rm ext}(0)=\ket{\Psi(0)}\bra{\Psi(0)}$ with
\begin{equation}
    \ket{\Psi(0)}=\frac{1}{\sqrt{2}}\qty(\ket{\xi^{(1)}(0)}+\ket{\xi^{(2)}(0)}).
\end{equation}

Let us say that the superposition state lasts for a time in the interval $0<t<t_f$ and $\xi^{(1)}(t)=\xi^{(2)}(t)$ holds for $t\notin [0,t_f]$. This means that the decoherence turned both trajectories indistinguishable. We can safely say that the coherence between the two components of this state will be effectively lost when $\exp\qty[-\Gamma(t_f)]<<1$ ---since the decay is exponential as we saw in the last section---, where $\Gamma(t_f)$ is the decoherence rate~\cite{Kanno2021}. Such a rate can be estimated from Eq.~\eqref{rho(xi,xi',t)-2}, resulting in
\begin{equation}
\begin{split}
    \Gamma(t_f)=\delta^{ij}\delta^{kl}\int_0^{t_f}dtdt'\,V_i(t)\Delta v_j(t)N_{\rm int}(t,t') \\
    \times V_k(t')\Delta v_l(t') \\
    +2\int_0^{t_f}dtdt'\,\Xi_i(t)\Delta\xi_j(t)\qty[1+\frac{1}{m_0^2}2N_{\rm int}(t,t')] \\
    \times N^{ijkl}_{\rm g}(t,t')\Xi_k(t')\Delta\xi_l(t'),
\end{split}
\end{equation}
where
\begin{equation*}
    \Xi_i\equiv\frac{1}{2}\qty(\xi_i^{(1)}+\xi_i^{(2)}),\hspace{0.3cm}\Delta\xi_i\equiv\xi_i^{(1)}-\xi_i^{(2)},
\end{equation*}
and similarly for $V_i$ and $\Delta v_i$. Since the quantities $V_i$ and $\Xi$ are averages of the paths in superposition, we will assume these to be time independent~\cite{Kanno2021}. In Appendix~\ref{A:Internal-degrees-of-freedom-noise-kernel} we explicitly compute $N_{\rm int}$ by considering an Ohmic bath in the high temperature limit. The internal degrees of freedom noise kernel is given in Eq.~\eqref{Int-noise-high-T}. In this case, the decoherence rate becomes
\begin{equation} \label{Decoherence-rate}
\begin{split}
    \Gamma(t_f)&=\frac{\lambda^2\gamma\pi}{\beta}\delta^{ij}\delta^{kl}V_iV_k\int_0^{t_f}dt\,\Delta v_j(t)\Delta v_l(t) \\
    &\hspace{0.2cm}+2\Xi_i\Xi_k\int_0^{t_f}dtdt'\,\Delta\xi_j(t)N_{\rm g}^{ijkl}(t,t')\Delta\xi_l(t') \\
    &\hspace{0.2cm}+\frac{4\lambda^2\gamma\pi}{\beta m_0^2}\Xi_i\Xi_k\int_0^{t_f}dt\,\Delta\xi_j(t)N_{\rm g}^{ijkl}(t)\Delta\xi_l(t),
\end{split}
\end{equation}
where $N_{\rm g}^{ijkl}(t)\equiv\lim_{t'\to t}N_{\rm g}^{ijkl}(t,t')$ while $\gamma$ represents the effective coupling constant and $\beta$ is the inverse temperature associated with the internal degrees of freedom.

Let us now consider a specific configuration of the superposition state. Take the separation $\Delta\xi_i(t)=\xi_i^{(2)}(t)-\xi_i^{(1)}(t)$ to be given by \cite{Kanno2021,Breuer2001}
\begin{subequations}
\begin{equation}
    \Delta\xi_i(t)=\left\{
    \begin{array}{ll}
        2v_it &\textrm{for}\hspace{0.2cm}0<t\leq t_f/2\\
        2v_i(t_f-t) &\textrm{for}\hspace{0.2cm}t_f/2<t<t_f
    \end{array}
    \right.,
\end{equation}
such that
\begin{equation}
    \Delta v_i(t)=\left\{
    \begin{array}{ll}
        2v_i &\textrm{for}\hspace{0.2cm}0<t\leq t_f/2\\
        -2v_i &\textrm{for}\hspace{0.2cm}t_f/2<t<t_f
    \end{array}
    \right..
\end{equation}
\end{subequations}
For this specific configuration, the decoherence rate~\eqref{Decoherence-rate} turns out to be
\begin{equation}
\begin{split}   \Gamma(t_f)&=8\Xi_iv_j\Xi_kv_l\left[ \int_0^{t_f/2}dtdt'\,tt'N_{\rm g}^{ijkl}(t,t')\right. \\
    &\hspace{1.0cm}+\int_{t_f/2}^{t_f}dtdt'\,(t_f-t)(t_f-t')N_{\rm g}^{ijkl}(t,t') \\
    &\hspace{1.0cm}\left. +2\int_0^{t_f/2}dt\int_{t_f/2}^{t_f}dt'\,t(t_f-t')N_{\rm g}^{ijkl}(t,t')\right] \\
    &\hspace{0.2cm}+\frac{16\lambda^2\gamma\pi}{\beta m_0^2}\Xi_iv_j\Xi_kv_l\left[ \int_0^{t_f/2}dt\,t^2N_{\rm g}^{ijkl}(t)\right. \\
    &\hspace{1.0cm}\left.+\int_{t_f/2}^{t_f}dt\,(t_f-t)^2N_{\rm g}^{ijkl}(t)\right].
\end{split}
\end{equation}

In order to proceed, we need to specify the state of the gravitational field. We show here only the results of the calculations considering four distinct initial states: $i$) vacuum; $ii$) thermal; $iii$) coherent; $iv$) squeezed. The details of the calculations can be found in Appendix~\ref{A:Gravitational-noise-kernel}

%%%%%%%%%%%%%%%%%%%%%%%%%%%%%%%%%%%%%%%%%%%%%%%%%%
\subsubsection{Vacuum state}

For gravitons initially in the vacuum state, the noise kernel is given in Eq.~\eqref{Noise-kernel-Minkowski-vacuum}, from which we can compute the decoherence rate as
\begin{equation} \label{Decoherence-rate-Minkowski-vacuum}
    \Gamma_{\rm (vac)}(t_f)=\frac{8m_0^2}{5\pi}\Lambda_{\rm g}^2\mathcal{K}\qty[G(\Lambda_{\rm g}t_f)+\lambda^2\kappa\qty(\Lambda_{\rm g}t_f)^3],
\end{equation}
where
\begin{equation}
\begin{split}
    G(x)&\equiv1+\frac{2}{3x}\qty[\sin x-8\sin\qty(\frac{x}{2})] \\
    &\hspace{0.5cm}+\frac{1}{x^2}\qty[\frac{2}{3}\cos x-\frac{32}{3}\cos\qty(\frac{x}{2})+10].
\end{split}
\end{equation}
$\mathcal{K}\equiv\mathcal{P}^{ijkl}\Xi_iv_j\Xi_kv_l$, while
\begin{equation}
\kappa\equiv\frac{\gamma\pi}{108m_0^2}\frac{\Lambda_{\rm g}}{\beta}.
\end{equation}
$\mathcal{P}^{ijkl}$ is a function of the Kronecker delta explicitly given in Appendix \ref{A:Gravitational-noise-kernel} and $\Lambda_{\rm g}$ is an appropriate cutoff. We observe that when $\lambda=0$ (no internal degrees of freedom), we recover the same qualitative behavior found in Ref.~\cite{Kanno2021}, as expected.

\begin{figure}[!h]
    \centering
    \includegraphics[width=1.0\linewidth]{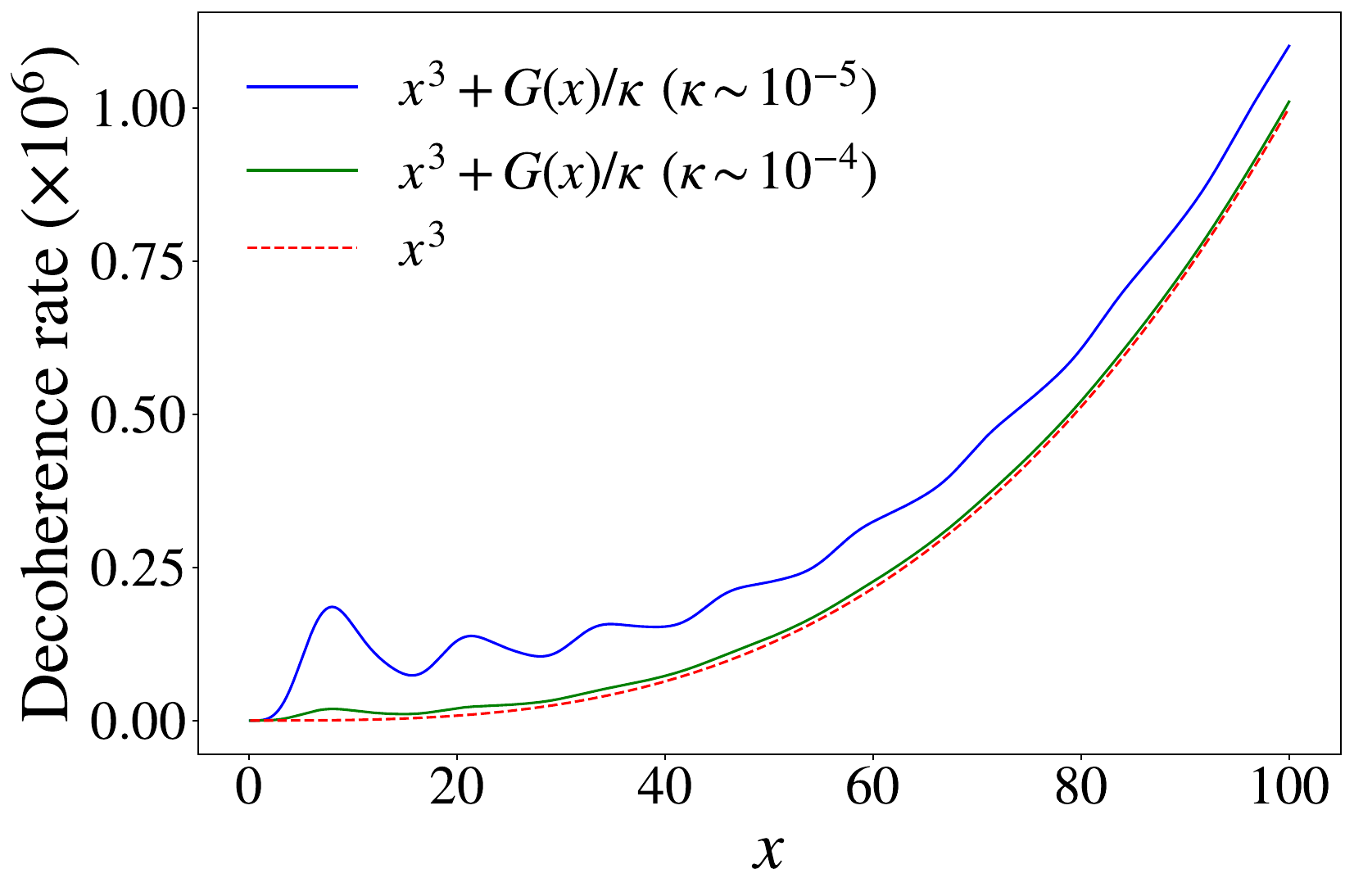}
    \caption{\textbf{Vacuum state}. Decoherence rate coming from the purely gravitational fluctuations and due to the presence of the internal degrees of freedom of our system. This last contribution becomes very important for $\kappa\gtrsim10^{-4}$.}
    \label{vac}
\end{figure}

The decoherence rate consists of the sum of two terms, one involving the decoherence induced exclusively by the quantum gravitational field and another one that comes from the coupling with both the internal degrees of freedom and the gravitons. The intensity of the latter term is given by a constant $\kappa$, which depends on the initial state considered. We expect that this term gives a greater contribution to the decoherence rate since it involves the coupling with two interacting environments and thus contains a further level of coarse-graining. In fact, Fig.~\ref{vac} shows how, even for small values of $\kappa$, the behavior of the decoherence rate is dominated by the term involving not only the gravitons but also the internal degrees of freedom of the particle.

%%%%%%%%%%%%%%%%%%%%%%%%%%%%%%%%%%%%%%%%%%%%%%
%%%%%%%%%%%%%%%%%%%%%%%%%%%%%%%%%%%%%%%%%%%%%%

The constant $\kappa$ quantifies the strength of the decoherence due to the presence of the internal degrees of freedom. Dropping the exclusively gravitational contribution, setting $\lambda=1$ and restoring the constants $\hbar$, $c$, $k_B$ and $G$, the contribution to the decoherence rate is given by
\begin{equation}
    \Gamma_{\rm (vac)}(t_f)\simeq\frac{8\mathcal{K}}{5\pi}\qty(\frac{m_0}{M_{\rm pl}})^2\kappa\Lambda_{\rm g}^2\qty(\Lambda_{\rm g}t_f)^3,
\end{equation}
where
\[
\kappa=\frac{1}{108}\varkappa\Lambda_{\rm g}T ,\hspace{1cm} \varkappa\equiv\frac{k_B\gamma\pi}{\hbar^3c^4m_0^2}
\]
while $M_{\rm pl}=\sqrt{\hbar c/G}$ is the Planck mass.

%%%%%%%%%%%%%%%%%%%%%%%%%%%%%%%%%%%%%%%%%%%%%%%
\subsubsection{Thermal state}

We now consider that the quantum gravitational field is in a thermal state with inverse temperature $\beta_{\rm g}^{-1}$. Before computing the kernel, it is important to observe that what we call temperature of gravitons may not be interpreted as temperature in a formal sense, but rather as a phenomenological parameter characterizing the power spectrum of the noise that is not related to any thermodynamic definition. It is important to keep this distinction in mind since gravitons interact very weakly and their thermalization cannot be assumed~\cite{Anastopoulos_2013}.

That being sad, the noise kernel is given in Eq.~\eqref{Noise-kernel-thermal-state}, thus resulting in the following expression for the decoherence rate
\begin{equation}
\begin{split}
    &\Gamma_{\rm (th)}(t_f)=\Gamma_{\rm (vac)}(t_f) \\
    &\hspace{0.2cm}+\frac{16m_0^2\pi}{15\beta_{\rm g}^2}\mathcal{K}\,\qty[G_{\rm th}\qty(\frac{\pi t_f}{\beta_{\rm g}})+\lambda^2\kappa_{\rm th}\qty(\frac{\pi t_f}{\beta_{\rm g}})^3],
\end{split}
\end{equation}
where
\begin{subequations}
\begin{equation}
    G_{\rm th}(x)\equiv\frac{1+16e^x+26e^{2x}+16e^{3x}+e^{4x}}{(e^{2x}-1)^2}-\frac{15}{x^2},
\end{equation}
and
\begin{equation}
    \kappa_{\rm th}\equiv\frac{4\gamma\pi^2}{189m_0^2\beta\beta_{\rm g}}.
\end{equation}
\end{subequations}

\begin{figure}[!h]
    \centering
    \includegraphics[width=1.0\linewidth]{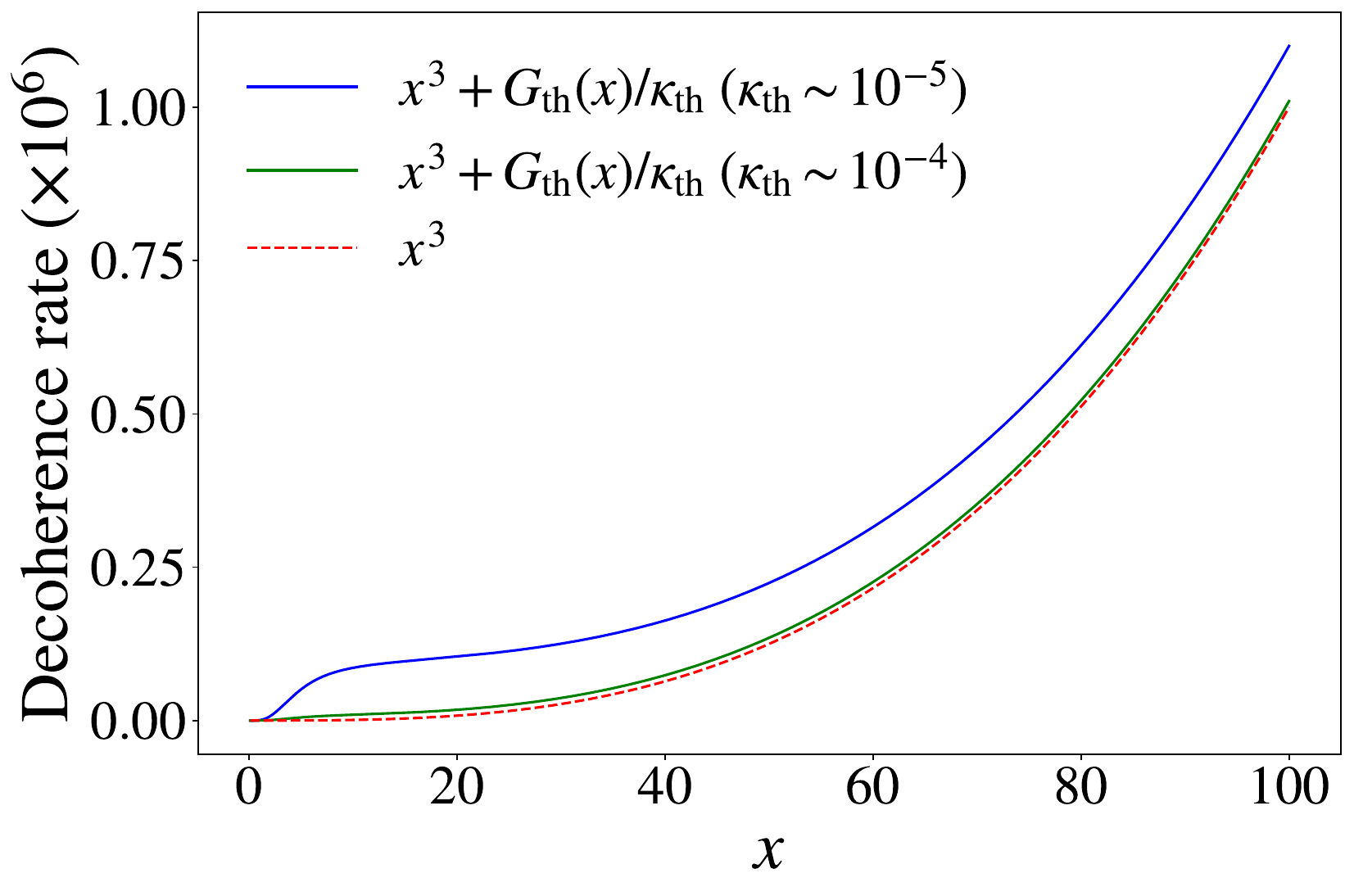}
    \caption{\textbf{Thermal state}. Decoherence rate coming from the purely gravitational fluctuations and due to the presence of the internal degrees of freedom of our system. This last contribution becomes very important for $\kappa_{\rm th}\gtrsim10^{-4}$.}
    \label{th}
\end{figure}

Figure~\ref{th} shows the comparison between the distinct contributions for the decoherence rate coming from the gravitational field and from the internal degrees of freedom.
%%%%%%%%%%%%%%%%%%%%%%%%%%%%%%%%%%%%%%%%%%%
%%%%%%%%%%%%%%%%%%%%%%%%%%%%%%%%%%%%%%%%%%

Although we observe a distinct quantitative behaviour with respect to the vacuum case, the qualitative pattern is the same, with the internal variables contributing significantly to the decoherence phenomenon. The intensity of such contribution is obtained by dropping the purely gravitational effect, setting $\lambda=1$ and restoring all the physical quantities, resulting in the decoherence rate
\begin{equation}
\begin{split}
    \Gamma_{\rm (th)}(t_f)&\simeq\Gamma_{\rm (vac)}(t_f) \\
    &+\frac{16\pi^4\mathcal{K}}{15}\qty(\frac{m_0}{M_{\rm pl}})^2\kappa_{\rm th}\qty(k_BT_{\rm g})^2\qty(k_BT_{\rm g}t_f)^3,
\end{split}
\end{equation}
with
\[
\kappa_{\rm th}=\frac{4\pi}{189}\varkappa k_BT_{\rm g}T
\]

%%%%%%%%%%%%%%%%%%%%%%%%%%%%%%%%%%%%%%%%%%%%
\subsubsection{Coherent state}

For gravitons initially in a coherent sate characterized by the real displacement parameter $\alpha$, the noise kernel can be found in Eq.~\eqref{Noise-kernel-coherent-state}, implying the decoherence rate
\begin{equation}
\begin{split}
    &\Gamma_{\rm (coh)}(t_f)=\Gamma_{\rm (vac)}(t_f) \\
    &\hspace{0.2cm}+\frac{128m_0^2\alpha^2\Tilde{\Lambda}_{\rm g}^2}{15\pi}\mathcal{K}\qty[G_{\rm coh}^{(I)}(\Tilde{\Lambda}_{\rm g}t_f)+\lambda^2\kappa_{\rm coh}G_{\rm coh}^{(II)}(\Tilde{\Lambda}_{\rm g}t_f)],
\end{split}
\end{equation}
where
\begin{subequations}
\begin{equation}
\begin{split}
    G_{\rm coh}^{(I)}(x)\equiv\frac{1}{1152x^2}\left\{ 1495+126x^2-1728\cos\qty(\frac{x}{2})\right. \\
    +288\cos x-64\cos\qty(\frac{3x}{2})+9\cos(2x)+18x\sin(2x) \\
    \left. -96x\qty[9\sin\qty(\frac{x}{2})-3\sin x+\sin\qty(\frac{3x}{2})]\right\} ,
\end{split}
\end{equation}
\begin{equation}
\begin{split}
    G_{\rm coh}^{(II)}(x)\equiv\frac{1}{12x^3}\left[ 441+2x^6+216\qty(x^2-2)\cos x\right. \\
    \left. +9\qty(2x^2-1)\cos(2x)-36x\qty(12-2x^2+\cos x)\sin x\right] ,
\end{split}
\end{equation}
and
\begin{equation}
    \kappa_{\rm coh}\equiv\frac{\gamma\pi\Tilde{\Lambda}_{\rm g}}{192m_0^2\beta}.
\end{equation}
\end{subequations}

\begin{figure}[!h]
    \centering
    \includegraphics[width=1.0\linewidth]{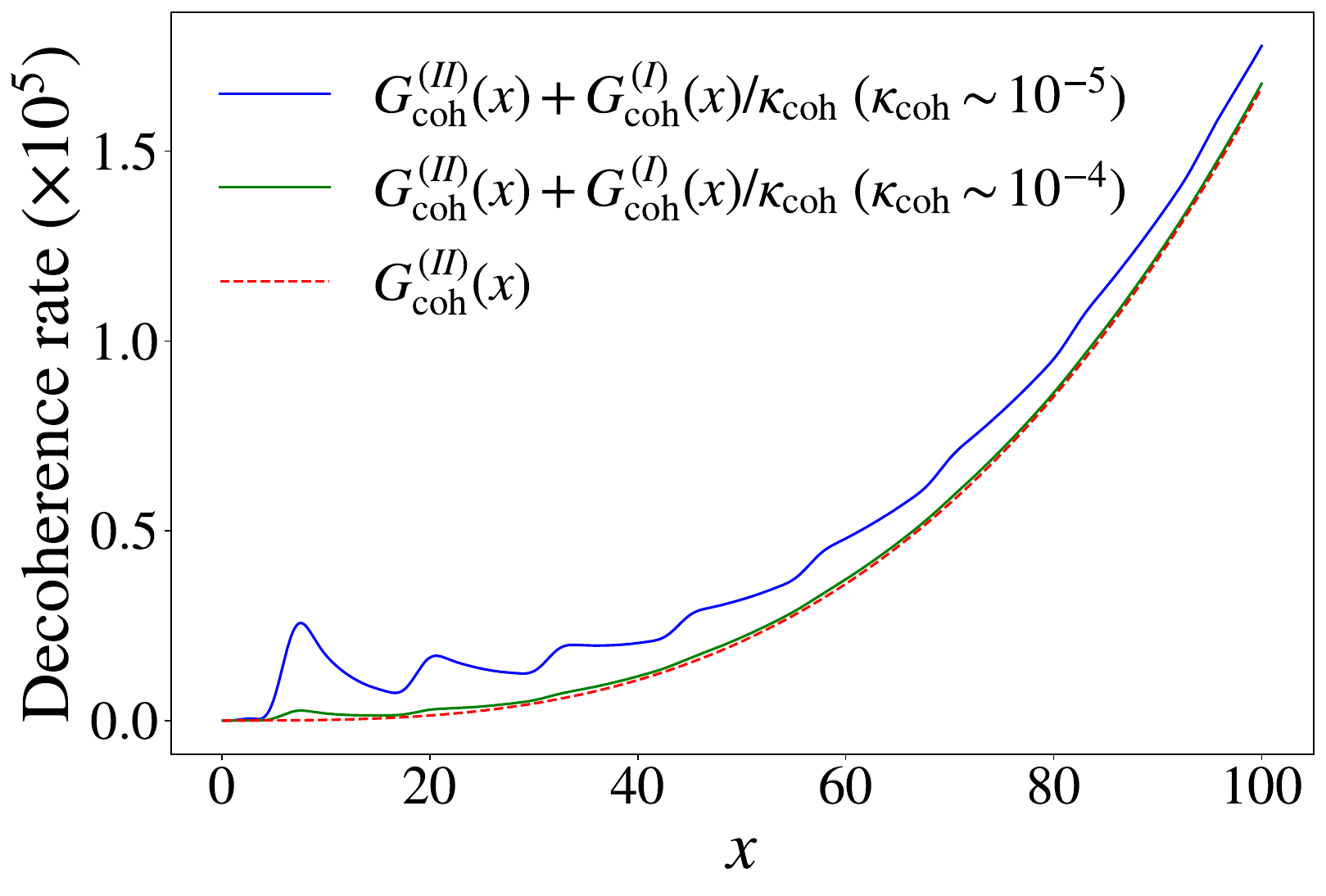}
    \caption{\textbf{Coherent state}. Decoherence rate coming from the purely gravitational fluctuations and due to the presence of the internal degrees of freedom of our system. This last contribution becomes very important for $\kappa_{\rm coh}\gtrsim10^{-4}$.}
    \label{coh}
\end{figure}

We again observe the same pattern that is illustrated in Fig.~\ref{coh}.
%%%%%%%%%%%%%%%%%%%%%%%%%%%%%%%%%%%%%%%%%%%
%%%%%%%%%%%%%%%%%%%%%%%%%%%%%%%%%%%%%%%%%%

For this case, the contribution to the decoherence rate coming from the internal degrees of freedom takes the form
\begin{equation}
\begin{split}
    \Gamma_{\rm (coh)}(t_f)&\simeq\Gamma_{\rm (vac)}(t_f) \\
    &+\frac{128\alpha^2\mathcal{K}}{15\pi}\qty(\frac{m_0}{M_{\rm pl}})^2\kappa_{\rm coh}\Tilde{\Lambda}_{\rm g}^2G_{\rm coh}^{(II)}(\Tilde{\Lambda}_{\rm g}t_f),
\end{split}
\end{equation}
with 
\[
\kappa_{\rm coh}=\frac{1}{192}\varkappa\Tilde{\Lambda}_{\rm g}T.
\]

We now move to our last example, the squeezed state.

%%%%%%%%%%%%%%%%%%%%%%%%%%%%%%%%%%%%%%%%
\subsubsection{Squeezed state}

Considering that the gravitons are initially in a squeezed sate parametrized by $\zeta=re^{i\varphi}$, we obtain the noise kernel shown in Eq.~\eqref{Noise-kernel-squeezed-state}. From this we directly obtain the decoherence rate
\begin{equation}
\begin{split}
    &\Gamma_{\rm (sq)}(t_f)=\cosh2r\,\Gamma_{\rm (vac)}(t_f) \\
    &\hspace{0.2cm}-\frac{2m_0^2\Bar{\Lambda}_{\rm g}^2}{135\pi}\mathcal{K}\sinh2r \\
    &\hspace{0.5cm}\times\qty[G_{\rm sq}^{(I)}(\Bar{\Lambda}_{\rm g}t_f;\varphi)+\lambda^2\kappa_{\rm sq}G_{\rm sq}^{(II)}(\Bar{\Lambda}_{\rm g}t_f;\varphi)],
\end{split}
\end{equation}
where
\begin{subequations}
\begin{equation}
\begin{split}
    G_{\rm sq}^{(I)}(x;\varphi)\equiv\frac{1}{x^2}\left[ -576\cos\qty(\frac{x}{2}-\varphi)+216\cos(x-\varphi)\right. \\
    -64\cos\qty(\frac{3x}{2}-\varphi)+9\cos(2x-\varphi) \\
    +(18x^2+415)\cos\varphi \\
    -288x\sin\qty(\frac{x}{2}-\varphi)+216x\sin(x-\varphi) \\
    \left. -96x\sin\qty(\frac{3x}{2}-\varphi)+18x\sin(2x-\varphi)\right] ,
\end{split}
\end{equation}
\begin{equation}
\begin{split}
    G_{\rm sq}^{(II)}(x;\varphi)\equiv\frac{1}{x^3}\left[ 72(x^2-2)\cos(x-\varphi)\right. \\
    +(6x^2-3)\cos(2x-\varphi)+147\cos\varphi \\
    -144x\sin(x-\varphi)+24x^3\sin(x-\varphi) \\
    \left. -6x\sin(2x-\varphi)-4x^3\sin\varphi\right] ,
\end{split}
\end{equation}
and
\begin{equation}
    \kappa_{\rm sq}\equiv\frac{3\gamma\pi\Bar{\Lambda}_{\rm g}}{2\beta m_0^2}.
\end{equation}
\end{subequations}

\begin{figure}[!h]
    \centering
    \includegraphics[width=1.0\linewidth]{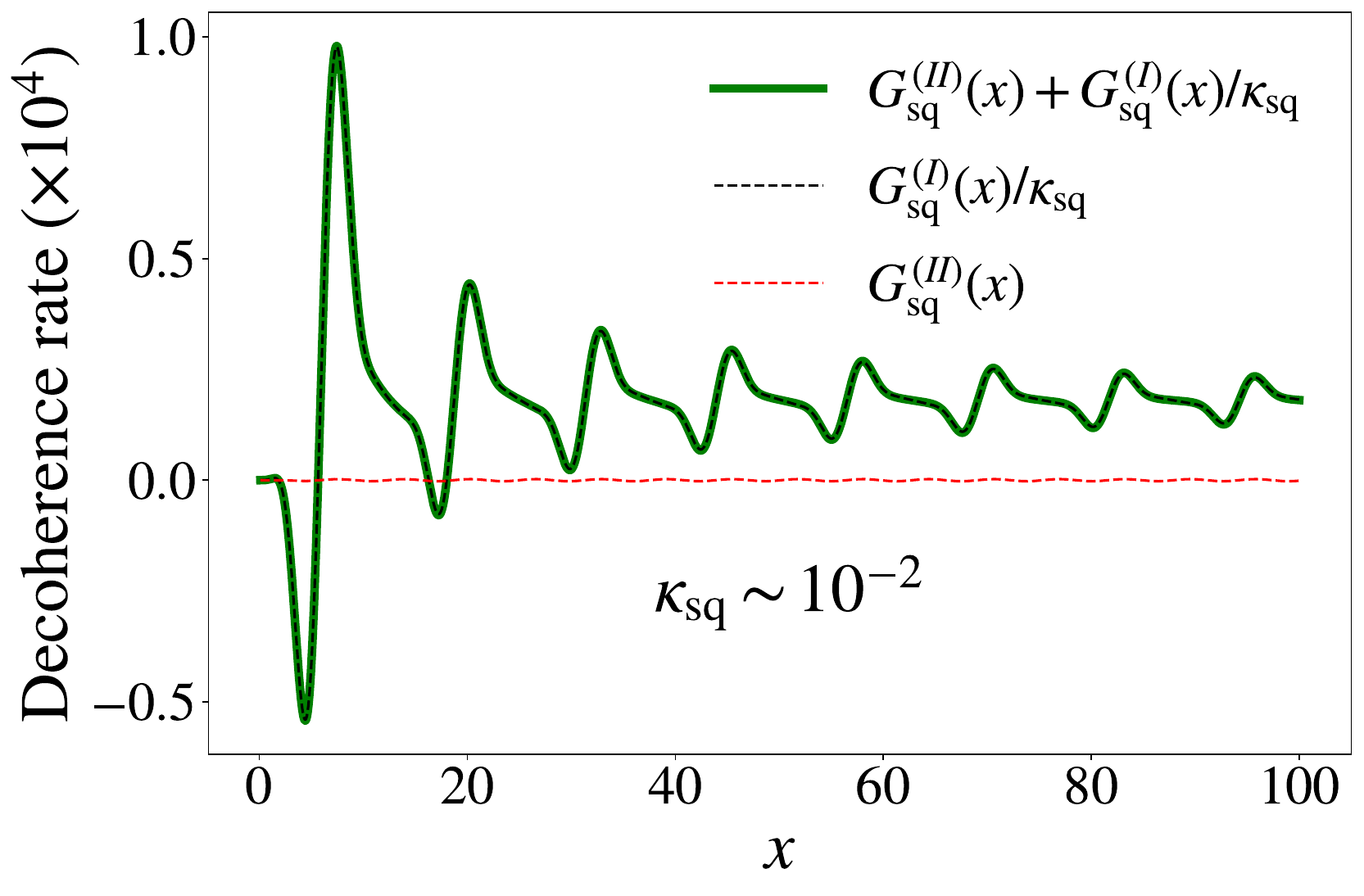}
    \includegraphics[width=1.0\linewidth]{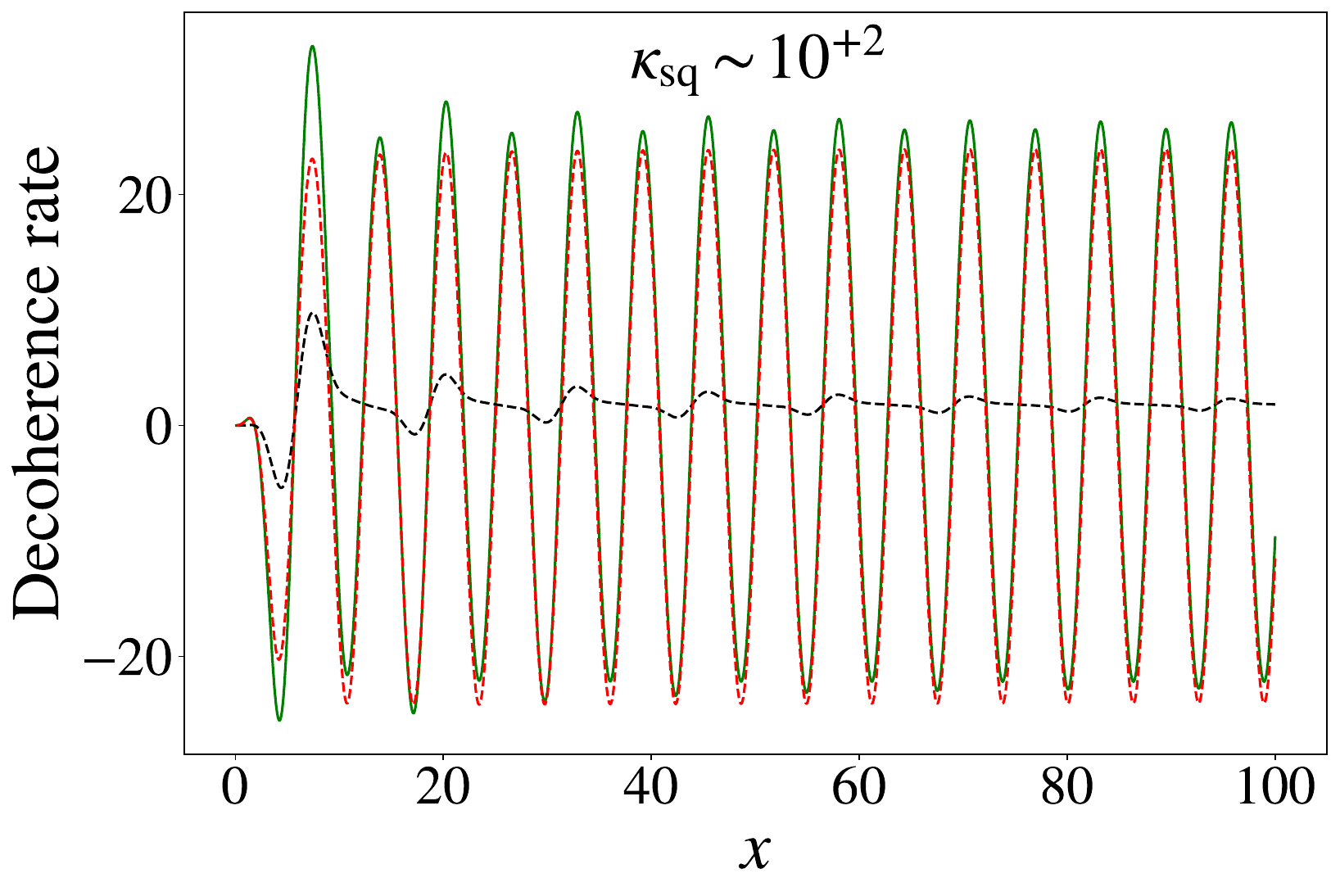}
    \caption{\textbf{Squeezed state}. The decoherence rate is found to be dominated by the corrections involving both gravitons and internal degrees of freedom for $\kappa_{\rm sq}\gtrsim10^{2}$, in contrast with the cases considered before where this correction dominates already for $\kappa\gtrsim10^{-4}$. In fact, even for $\kappa_{\rm sq}\sim10^{-2}$ the behavior of the decoherence rate is completely dominated by the exclusive graviton contribution.}
    \label{sq}
\end{figure}

Figure~\ref{sq} illustrates the dynamical behaviour of the decoherence rate. Although we obtained the same pattern here, in this case the contribution coming from purely gravitational fluctuations are dominant, except for very large interaction strength $\kappa$.

%%%%%%%%%%%%%%%%%%%%%%%%%%%%%%%%%%%%%%%%%%%%
%%%%%%%%%%%%%%%%%%%%%%%%%%%%%%%%%%%%%%%%%%%

Disregarding the contributions coming solely from the quantum fluctuations of the gravitational field, we end up with the following expression for the decoherence rate
\begin{equation}
\begin{split}
    \Gamma_{\rm (sq)}(t_f)&\simeq\cosh2r\,\Gamma_{\rm (vac)}(t_f) \\
    &-\frac{2\mathcal{K}}{135\pi}\sinh2r\qty(\frac{m_0}{M_{\rm pl}})^2\kappa_{\rm sq}\Bar{\Lambda}_{\rm g}^2G_{\rm sq}^{(II)}(\Bar{\Lambda}_{\rm g}t_f;\varphi),
\end{split}
\end{equation}
with
\[
\kappa_{\rm sq}=\frac{3}{2}\varkappa\Bar{\Lambda}_{\rm g}T.
\]

Since we have the decoherence rate, it is straight forward to obtain the decoherence time, which we present in the sequence for all the considered initial states.

%%%%%%%%%%%%%%%%%%%%%%%%%%%%%%%%%%%%%%%%%%%%%%%%%%%%
\subsubsection{Decoherence time}

As mentioned before, the decoherence time is obtained by setting $\Gamma(\tau_{\rm dec})=1$. The case of the vacuum and the thermal initial states of the field directly leads to the decoherence times
\begin{equation}
    \tau_{\rm dec}^{(\textrm{vac})}=\qty(\frac{5\pi}{8\mathcal{K}})^{1/3}\qty(\frac{M_{\rm pl}}{m_0})^{2/3}\qty(\frac{1}{\kappa\Lambda_{\rm g}^5})^{1/3},
\end{equation}
and
\begin{equation}
\begin{split}
    \tau_{\rm dec}^{(\textrm{th})}&=\qty(\frac{15\pi}{8\mathcal{K}})^{1/3}\qty(\frac{M_{\rm pl}}{m_0})^{2/3} \\
    &\hspace{1cm}\times\qty(\frac{1}{3\kappa\Lambda_{\rm g}^5+2\pi^5\kappa_{\rm th}(k_BT_{\rm g})^5})^{1/3},
\end{split}
\end{equation}
respectively. 

However, for the condition $\Gamma(\tau_{\rm dec})=1$ to hold we must have both $G_{\rm coh}^{(II)}$ and $G_{\rm sq}^{(II)}$ less than unit for the cases of the coherent and the squeezed states, respectively~\cite{Kanno2021}. Thus we use $G_{\rm coh}^{(II)}\simeq x^3/3$ and $G_{\rm sq}^{(II)}\simeq2x^3/3$, which hold for $x<<1$. Under such assumptions, we find the decoherence times for the coherent and the squeezed initial states to be given by
\begin{equation}
\begin{split}
    \tau_{\rm dec}^{(\textrm{coh})}&=\qty(\frac{45\pi}{8\mathcal{K}})^{1/3}\qty(\frac{M_{\rm pl}}{m_0})^{2/3} \\
    &\hspace{1cm}\times\qty(\frac{1}{9\kappa\Lambda_{\rm g}^5+16\alpha^2\kappa_{\rm coh}\Tilde{\Lambda}_{\rm g}^5})^{1/3},
\end{split}
\end{equation}
and
\begin{equation}
\begin{split}
    \tau_{\rm dec}^{(\textrm{sq})}&=\qty(\frac{405\pi}{4\mathcal{K}})^{1/3}\qty(\frac{M_{\rm pl}}{m_0})^{2/3} \\
    &\hspace{1cm}\times\qty(\frac{1}{162\kappa\Lambda_{\rm g}^5\cosh2r-\kappa_{\rm sq}\Bar{\Lambda}_{\rm g}^5\sinh2r})^{1/3},
\end{split}
\end{equation}
respectively.

As already pointed out by the authors of Ref.~\cite{Kanno2021}, the decoherence time is found to be proportional to the ratio $M_{\rm pl}/m_0$. For a point particle, decoherence does not occur for $m_0<<M_{\rm pl}$. However, when there are internal degrees of freedom the decoherence time is found to be reduced with respect to the one estimated in Ref.~\cite{Kanno2021}. Specifically when these degrees of freedom are in thermal equilibrium with a high temperature, the constants $\kappa$ are very large and the decoherence time becomes even smaller.

%%%%%%%%%%%%%%%%%%%%%%%%%%%%%%%%%%%%%%%%%%%%%%
\section{Conclusions} \label{S:Conclusions}

We considered the evolution of the external degrees of freedom of a freely falling particle under the action of two sources of noise: $i$) the coupling with its internal degrees of freedom and $ii$) the coupling with the quantized weak gravitational field. Due to the universal nature of the gravitational coupling, both environments are not independent. Employing the Feynman-Vernon influence functional technique we computed the decoherence rate considering a superposition of the relevant degrees of freedom. From this we estimated the decoherence time.

For all the cases considered here, the decoherence rate was found to be given by the sum of two terms. One arising purely from the quantum fluctuations of the gravitational field and another contribution that comes from the coupling with both the internal degrees of freedom and the gravitons, whose intensity depends on the initial state of the gravitational field. In order to reach our conclusions, we assumed that the temperature associated with the internal degrees of freedom is high enough so they can be treated as a white noise. Within this range of intensities analyzed here, we found that the presence of the dynamical internal degrees of freedom, which couples to the external variables due to the (quantum) gravitational field, causes a reduction in the decoherence time with respect to the one estimated in the literature by considering only the coupling with the gravitons. 

As described here, the corrections involving both of these interacting environments (gravitational field and internal dynamics) dominate except for the squeezed initial state of the gravitons. This indicates that gravitons in a squeezed state yields a much higher contribution to decoherence. Nevertheless, we expected that the intensity of the contribution coming from the internal dynamics to be much larger than the values considered here since they are proportional to the gravitational cutoff $\Lambda$ (or graviton temperature $T_{\rm g}=1/\beta_{\rm g}$ in the case of thermal initial state) as well as to the temperature of the internal degrees of freedom $T=1/\beta$. This will turn such contribution even more important.

Our results show that it is fundamental to take into account the quantum nature of matter as well as the quantum fluctuations of the gravitational field when considering phenomena like entanglement induced by gravity, quantum reference frames or decoherence induced by the gravitational time-dilation. Actually some of these effects can be completely blurred by the quantum fluctuations coming from the sources we discussed here. 

An interesting additional question that deserves investigation is the contribution from these fluctuations to the entropy production due to the universal coupling with gravity, as recently discussed in Ref.~\cite{Basso2023}.

\section*{Acknowledgements}
This work was supported by the the National Institute for the Science and Technology of Quantum Information (INCT-IQ), Grant No.~465469/2014-0, by the National Council for Scientific and Technological Development (CNPq), Grants No.~308065/2022-0 and Coordination of Superior Level Staff Improvement (CAPES).

\appendix

\section{Open quantum systems and the Feynman-Vernon influence functional} \label{A:Quantum-open-systems-and-the-Feynman-Vernon-influence-functional}

In this Appendix we present the formalism of the Feynman-Vernon influence functional in order to study the general dynamics of an open quantum system. We follow the presentation in Ref.~\cite{Calzetta2008}. Let us consider that the system of interest is described by the variables $x=\qty{x_n}$ and interacts with an environment described by the variables $q=\qty{q_n}$. The classical action takes the form
\begin{equation}
    S[x,q]=S_{\rm sys}[x]+S_{\rm env}[q]+S_{\rm int}[x,q],
\end{equation}
where the action $S_{\rm int}[x,q]$ describes the interaction between the system and the environment. The quantum state of the total system is described by the density matrix with matrix elements $\rho(xq,x'q',t)=\mel{xq,t}{\rho}{x'q,t}$ depending on both system and environment variables that evolves unitarily under the total Hamiltonian. However, we are only interested in computing the expectation values of system observables without taking the environment specific quantum state into account. The expectation value of such observables may be computed with the reduced density matrix $\rho_r=\textrm{Tr}_q\,\rho$.

Assuming that at $t=0$ the system and environment are uncorrelated,
\begin{equation}
    \rho(x_iq_i,x_i'q_i',0)=\rho_{\rm sys}(x_ix_i',0)\rho_{\rm env}(q_iq_i',0),
\end{equation}
and using the path integral representation of transition amplitudes, the reduced density matrix evolves in time according to
\begin{equation} \label{Evolution-of-reduced-density-matrix}
    \rho_r(xx',t)=\int dx_idx_i'\,\mathcal{J}_r(xx',t|x_ix_i',0)\rho_{\rm sys}(x_ix_i',0),
\end{equation}
where the evolution operator for the reduced density matrix is defined by
\begin{equation}
\begin{split}
    \mathcal{J}_r(xx',t|x_ix_i',0)&\equiv\int_{x_i}^x\mathcal{D}x\int_{x_i'}^{x'}\mathcal{D}x'\, \\
    &\hspace{0.5cm}\times e^{i\qty(S_{\rm sys}[x]-S_{\rm sys}[x'])}\mathcal{F}[x,x'].
\end{split}
\end{equation}
The functional $\mathcal{F}[x,x']$ is the so-called Feynman-Vernon influence functional~\cite{Calzetta2008,Feynman1963,Feynman2010}, and it is given by
\begin{align} \label{Influence-functional-1}
    \mathcal{F}[x,x']&=e^{iS_{\rm IF}[x,x',t]} \nonumber \\
    &=\int_{-\infty}^\infty dqdq_idq_i'\,\rho_{\rm env}(q_iq_i',0) \nonumber \\
    &\hspace{0.5cm}\times\int_{q_i}^q\mathcal{D}q\,e^{i\qty(S_{\rm env}[q]+S_{\rm int}[x,q])} \nonumber \\
    &\hspace{0.5cm}\times\int_{q_i'}^q\mathcal{D}q'\,e^{-i\qty(S_{\rm env}[q']+S_{\rm int}[x',q'])},
\end{align}
where $S_{\rm IF}$ is called the influence action and it encodes all the influence of the environment on the system.

The path integral representation \eqref{Influence-functional-1} involves two histories, which may be described as an integral over single histories defined on a \emph{closed time path} (CTP).
%This time path has a first branch from $0$ to $t$, where the history takes the values $x(t)$, and a second branch from $t$ back to $0$, where the history takes the values $x'(t)$, and the CTP boundary condition reads $x(t)=x'(t)$ \cite{Calzetta2008}.
For the remainder of this appendix, we denote $x(t)=x^1(t)$, $x'(t)=x^2(t)$ and we shall think of $x^a$, $a=1,2$, as a single doublet field defined on a single branch time path. We also define a metric tensor $c_{ab}=\textrm{diag}\qty(1,-1)$ that, together with its inverse $c^{ab}=\textrm{diag}\qty(1,-1)$ (defined such that $c^{ad}c_{db}=\delta^a_b$), may be used to raise and/or lower indices.
%, as in \cite{Calzetta2008}
%\begin{equation*}
%    \begin{split}
%        x_1&=c_{1a}x^a=c_{11}x^1=x^1=x, \\
%        x_2&=c_{2a}x^a=c_{22}x^2=-x^2=-x'.
%    \end{split}
%\end{equation*}
Using the Einstein summation convention, the kinetic terms in the system action will be written as
\begin{equation*}
    \frac{1}{2}\int dt\,c_{ab}\dot{x}^a\dot{x}^b=\frac{1}{2}\int dt\,\dot{x}_a\dot{x}^a=\frac{1}{2}\int dt\,\qty(\dot{x}^2-\dot{x}'^2).
\end{equation*}
We will refer to the \emph{CTP action} $S[x^a]\equiv S[x]-S[x']$ without discriminating the contributions from either branch \cite{Calzetta2008}.

\subsection{The linear coupling model} \label{S:The-linear-coupling-model}

Let us consider the case where the environment action is quadratic in the $q$ variable(s), the initial environment density matrix is Gaussian and the interaction term is bilinear,
\begin{equation} \label{Linear-bath-action}
    S_{\rm int}=\int dt\,x^a(t)Q_a[q(t)],
\end{equation}
where the $Q$'s are linear combinations of the $q$'s \cite{Calzetta2008}. Note that, in that case, Eq. \eqref{Influence-functional-1} is a functional Fourier transform of a Gaussian functional of histories $Q(t)$ and $Q'(t)$. Since the Fourier transform of a Gaussian is another Gaussian, we conclude that, under all these assumptions, the influence action must also be quadratic in $x$ and $x'$. Therefore we write
\begin{equation}
    S_{\textrm{IF}}=\frac{1}{2}\int dtdt'\,x^a(t)\mathbb{M}_{ab}(t,t')x^b(t'),
\end{equation}
where
\begin{equation}
    \mathbb{M}_{ab}(t,t')=-i\frac{\delta^2}{\delta x^a(t)\delta x^b(t')}\eval{e^{iS_{\rm IF}[x^a]}}_{x_a=0}.
\end{equation}

Eq. \eqref{Influence-functional-1} in CPT notation reads
\begin{equation*}
\begin{split}
    \mathcal{F}[x^a]&=\int_{-\infty}^\infty dq^1(t)dq^1(0)dq^2(0) \\
    &\times\int\mathcal{D}q^a\,e^{i\qty(S_{\rm env}[q^a]+S_{\rm int}[x^a,q^a])}\rho_{\rm env}(q^1(0),q^2(0),0),
\end{split}
\end{equation*}
where the path integral is taken under the CTP boundary condition $q^1(t)=q^2(t)$. A direct variation from this equation with $S_{\rm int}$ given by Eq. \eqref{Linear-bath-action} yields
\begin{widetext}
\begin{equation}
    \frac{\delta^2}{\delta x^a(t)\delta x^b(t')}\eval{e^{iS_{\rm IF}[x^a]}}_{x_a=0}=-\int_{-\infty}^\infty dq^1(t)dq^1(0)dq^2(0)\int\mathcal{D}q^a\,e^{iS_{\rm env}[q^a]}Q_a(t)Q_b(t')\rho_{\rm env}(q^1(0),q^2(0),0).
\end{equation}
\end{widetext}

After a long calculation \cite{Calzetta2008}, we find
\begin{equation}
    \mathbb{M}_{ab}(t,t')=i\mqty(\expval{T\qty[Q(t)Q(t')]} & -\expval{Q(t')Q(t)} \\ -\expval{Q(t)Q(t')} & \expval{\Tilde{T}\qty[Q(t)Q(t')]}),
\end{equation}
where $T$ stands for temporal ordering (the latest time to the left) and $\Tilde{T}$ stands for \emph{anti}-time ordering (the latest time to the right). Here, expectation values are understood to be taken with respect to the initial state of the environment. In this last expression, the $Q'$s are now understood as operators in the Heisenberg representation. Then, writing the influence action in terms of the original variables, and after another long calculation which involves only algebraic manipulations, we find

\begin{equation} \label{Influence-Action:Dissipation-and-Noise-Kernel}
\begin{split}
    S_{\rm IF}=\int dtdt'\,\left\{ \frac{1}{2}\qty[x(t)-x'(t)]D(t,t')\qty[x(t')+x'(t')]\right. \\
    \left. +\frac{i}{2}\qty[x(t)-x'(t)]N(t,t')\qty[x(t')-x'(t')]\right\} ,
\end{split}
\end{equation}
where we have defined the \emph{dissipation} $D$ and \emph{noise} $N$ kernels

\begin{subequations}
    \begin{equation}
        D(t,t')=i\expval{\comm{Q(t)}{Q(t')}}\theta(t-t'),
    \end{equation}
    \begin{equation}
        N(t,t')=\frac{1}{2}\expval{\acomm{Q(t)}{Q(t')}}.
    \end{equation}
\end{subequations}

\section{Internal degrees of freedom noise kernel} \label{A:Internal-degrees-of-freedom-noise-kernel}

The internal degrees of freedom noise kernel is given by Eq. \eqref{Internal-dofs-noise-kernel}. Since the free Lagrangian for each $\varrho_\alpha$ is the one of a harmonic oscillator, we may write the position operators in the Heisenberg picture as
\begin{equation*}
    \varrho_\alpha(t)=\sqrt{\frac{1}{2\mu_\alpha\varpi_\alpha}}\qty(b_\alpha e^{-i\varpi_\alpha t}+b_\alpha^\dagger e^{i\varpi_\alpha t}),
\end{equation*}
where the $b$'s ($b^\dagger$'s) are annihilation (creation) operators. A direct calculation then yields
\begin{equation}
\begin{split}
    \expval{\acomm{\varrho_\alpha(t)}{\varrho_\alpha(t')}}_{\rm int}=\frac{2}{\mu_\alpha\varpi_\alpha^2}\expval{H_\alpha}_{\rm int}\cos\varpi_\alpha(t-t') \\
    +\frac{1}{\mu_\alpha\varpi_\alpha}\qty[\expval{b_\alpha^2}_{\rm int}e^{-i\varpi_\alpha(t+t')}+\expval{\qty(b^\dagger_\alpha)^2}_{\rm int}e^{i\varpi_\alpha(t+t')}],
\end{split}
\end{equation}
where $H_\alpha$ is the free Hamiltonian operator of the $\alpha$th oscillator with frequency $\varpi_\alpha$.

Let us take the initial internal state to be a thermal one with temperature $1/\beta$, so that $\expval{b_\alpha^2}_{\rm int}=\expval{\qty(b^\dagger_\alpha)^2}_{\rm int}=0$. The canonical partition function reads
\begin{equation*}
    Z=\sum_{n=0}^\infty\exp\qty[-\varpi_\alpha\beta\qty(n+\frac{1}{2})]=\frac{e^{-\varpi_\alpha\beta/2}}{1-e^{-\varpi_\alpha\beta}},
\end{equation*}
and therefore
\begin{equation}
    \expval{H_\alpha}_{\rm int}=\frac{\varpi_\alpha}{2}\coth\qty(\frac{\varpi_\alpha\beta}{2}).
\end{equation}
We are then left with
\begin{equation}
    \expval{\acomm{\varrho_\alpha(t)}{\varrho_\alpha(t')}}_{\rm int}=\frac{1}{\mu_\alpha\varpi_\alpha}\coth\qty(\frac{\varpi_\alpha\beta}{2})\cos\varpi_\alpha(t-t'),
\end{equation}
and Eq. \eqref{Internal-dofs-noise-kernel} becomes
\begin{equation}
    N_{\rm int}(t,t')=\frac{\lambda^2}{2}\sum_\alpha\frac{\vartheta_\alpha^2}{\mu_\alpha\varpi_\alpha}\coth\qty(\frac{\varpi_\alpha\beta}{2})\cos\varpi_\alpha(t-t').
\end{equation}

Let us further assume that the internal frequencies span a continuum, so we may replace
\begin{equation}
    \sum_\alpha\to\int_0^\infty d\varpi\,\sigma(\varpi),
\end{equation}
where $\sigma(\varpi)d\varpi$ is the number of oscillators with frequencies between $\varpi$ and $\varpi+d\varpi$. Then,
\begin{equation}
    N_{\rm int}(t,t')=\frac{\lambda^2}{2}\int_0^\infty\frac{d\varpi}{\varpi}\Omega(\varpi)\coth\qty(\frac{\varpi\beta}{2})\cos\varpi(t-t'),
\end{equation}
with
\begin{equation*}
    \Omega(\varpi)\equiv\frac{\sigma(\varpi)\vartheta^2(\varpi)}{\mu_\varpi}.
\end{equation*}
If we consider the internal degrees of freedom to represent an ohmic bath, we have \cite{Calzetta2008}
\begin{equation}
    \Omega(\varpi)=\gamma\varpi^2
\end{equation}
for some constant $\gamma$. For that case, we have
\begin{equation} \label{Int-noise}
    N_{\rm int}(t,t')=\frac{1}{2}\lambda^2\gamma\int_0^\infty d\varpi\,\varpi\coth\qty(\frac{\varpi\beta}{2})\cos\varpi(t-t').
\end{equation}
At high temperature, $\beta<<t-t'$, the integral is dominated by low frequencies, and we find
\begin{equation} \label{Int-noise-high-T}
    N_{\rm int}(t,t')=\frac{\pi\lambda^2\gamma}{\beta}\delta(t-t'),
\end{equation}
namely a white noise.

\section{Gravitational noise kernel} \label{A:Gravitational-noise-kernel}

In this appendix we compute the noise kernel enconding the influence of the quantized gravitational field, defined by Eq. \eqref{Noise-kernel-definition}, for different initial states. In what follows, we shall consider the position operators to be independent of the polarizations and of the direction of $\vb{k}$, that is, $q_s(t,\vb{k})=q_\omega(t)$, $\omega=\abs{\vb{k}}$. This means that the angular integral we need to compute is
\begin{equation*}
    \int d\Omega\sum_s\epsilon_s^{ij}(\vb{k})\epsilon_s^{kl}(\vb{k}).
\end{equation*}

The polarization tensors can be written as \cite{Carroll,Cho2022}
\begin{equation}
\begin{split}
    \epsilon_{ij}^{+}&=\hat{\epsilon}_i^{(1)}\hat{\epsilon}_j^{(1)}-\hat{\epsilon}_i^{(2)}\hat{\epsilon}_j^{(2)} \\
    \epsilon_{ij}^{\cross}&=\hat{\epsilon}_i^{(1)}\hat{\epsilon}_j^{(2)}+\hat{\epsilon}_i^{(2)}\hat{\epsilon}_j^{(1)},
\end{split}
\end{equation}
where the spatial polarization unit vectors $\hat{\epsilon}^{(1)}$ and $\hat{\epsilon}^{(2)}$ are orthogonal to the direction of propagation $\hat{k}=\vb{k}/\abs{\vb{k}}$. They satisfy
\begin{equation}
    \hat{\epsilon}_i^{(1)}\hat{\epsilon}_j^{(1)}+\hat{\epsilon}_i^{(2)}\hat{\epsilon}_j^{(2)}=\delta_{ij}-\hat{k}_i\hat{k}_j\equiv P_{ij}.
\end{equation}
Then, a straightforward calculation yields
\begin{equation}
    \sum_s\epsilon_s^{ij}\epsilon_s^{kl}=P^{ik}P^{jl}+P^{il}P^{jk}-P^{ij}P^{kl}.
\end{equation}

We can use the following integration over solid angles:
\begin{equation*}
    \int d\Omega\,\hat{k}_i\hat{k}_j=\frac{4\pi}{3}\delta_{ij},
\end{equation*}
\begin{equation*}
    \int d\Omega\,\hat{k}_i\hat{k}_j\hat{k}_k\hat{k}_l=\frac{4\pi}{15}\qty(\delta_{ij}\delta_{kl}+\delta_{ik}\delta_{jl}+\delta_{il}\delta_{jk}),
\end{equation*}
to obtain
\begin{equation*}
    \int d\Omega P_{ij}P_{kl}=\frac{8\pi}{5}\delta_{ij}\delta_{kl}+\frac{4\pi}{15}\qty(\delta_{ij}\delta_{jl}+\delta_{il}\delta_{jk}).
\end{equation*}

We finally obtain
\begin{subequations}
\begin{equation}
    \int d\Omega\sum_s\epsilon_s^{ij}(\vb{k})\epsilon_s^{kl}(\vb{k})=\frac{8\pi}{15}\mathcal{P}^{ijkl},
\end{equation}
where
\begin{equation}
    \mathcal{P}^{ijkl}\equiv3\qty(\delta^{ik}\delta^{jl}+\delta^{il}\delta^{jk})-2\delta^{ij}\delta^{kl}.
\end{equation}
\end{subequations}

In order to compute the noise kernel \eqref{Noise-kernel-definition}, we will need to evaluate the expectation value of the anti-commutator of position operators at different times with respect to the gravitons initial state. Since the free Lagrangian for each mode of the gravitational field is the one of a harmonic oscillator, we may write the position operators in the Heisenberg picture as
\begin{equation*}
    q_\omega(t)=\sqrt{\frac{1}{2m\omega}}\qty(a_\omega e^{-i\omega t}+a_\omega^\dagger e^{i\omega t}),
\end{equation*}
where the $a$'s ($a^\dagger$'s) are annihilation (creation) operators satisfying the usual commutation relations. A direct calculation then yields
\begin{equation} \label{Expval-anticomm}
\begin{split}
    &\expval{\acomm{q_\omega(t)}{q_\omega(t')}}_{\rm g}=\frac{2}{m\omega^2}\expval{H_\omega}_{\rm g}\cos\omega(t-t') \\
    &\hspace{0.2cm}+\frac{1}{m\omega}\qty[\expval{a_\omega^2}_{\rm g}e^{-i\omega(t+t')}+\expval{\qty(a^\dagger_\omega)^2}_{\rm g}e^{i\omega(t+t')}]
\end{split}
\end{equation}
where
\begin{equation*}
    H_\omega=\omega\qty(a_\omega^\dagger a_\omega+\frac{1}{2})=\frac{\omega}{2}\acomm{a_\omega}{a_\omega^\dagger}
\end{equation*}
is the free Hamiltonian operator of the oscillator with frequency $\omega$. We may now compute the noise kernel for different initial states of the gravitational field.

\subsection{Minkowski vacuum}

If the initial state is the vacuum state, we have $\expval{a_\omega^2}_{\rm g}=\expval{\qty(a^\dagger_\omega)^2}_{\rm g}=0$ and $\expval{H_\omega}_{\rm g}=\omega/2$, so that
\begin{equation}
    \expval{\acomm{q_\omega(t)}{q_\omega(t')}}_{\rm g}=\frac{1}{m\omega}\cos\omega(t-t').
\end{equation}

Then, from Eq. \eqref{Noise-kernel-definition} we have
\begin{equation*}
\begin{split}
    N_{\textrm{g (vac)}}^{ijkl}(t,t')&=\frac{m_0^2}{15\pi}\mathcal{P}^{ijkl}\int_0^\infty d\omega\,\omega^5\cos\omega(t-t').
\end{split}
\end{equation*}
Since the integral over $\omega$ is divergent in its upper limit, we need to introduce the momentum cutoff $\Lambda_{\rm g}$, so that \cite{Cho2022,Kanno2021}
\begin{subequations}
\begin{equation}
    \int_0^{\Lambda_{\rm g}}d\omega\,\omega^5\cos\omega(t-t')=\Lambda_{\rm g}^6\,F\qty[\Lambda_{\rm g}(t-t')],
\end{equation}
where
\begin{align} \label{F(x)}
    F(x)&\equiv\frac{1}{x^6}\int_0^xdy\,y^5\cos y \nonumber \\
    &=\frac{1}{x^6}\left[ \qty(5x^4-60x^2+120)\cos x\right. \nonumber \\ 
    &\hspace{1.0cm}\left. +x\qty(x^4-20x^2+120)\sin x-120\right] .
\end{align}
\end{subequations}

Then, the noise kernel for the Minkowski vacuum initial state reads
\begin{equation} \label{Noise-kernel-Minkowski-vacuum}
    N_{\textrm{g (vac)}}^{ijkl}(t,t')=\frac{m_0^2\Lambda_{\rm g}^6}{15\pi}\mathcal{P}^{ijkl}F\qty[\Lambda_{\rm g}(t-t')].
\end{equation}

\subsection{Thermal state}

If the gravitons are initially in a thermal state with temperature $1/{\beta_{\textrm{g}}}$, we have $\expval{a_\omega^2}_{\rm g}=\expval{\qty(a^\dagger_\omega)^2}_{\rm g}=0$ once again. The canonical partition function reads
\begin{equation*}
    Z_{\rm g}=\sum_{n=0}^\infty\exp\qty[-\omega\beta_{\rm g}\qty(n+\frac{1}{2})]=\frac{e^{-\omega\beta_{\rm g}/2}}{1-e^{-\omega\beta_{\rm g}}},
\end{equation*}
from which we find
\begin{equation}
    \expval{H_\omega}_{\rm g}=-\pdv{\beta_{\rm g}}\ln Z_{\rm g}=\omega\qty(\frac{1}{2}+\frac{1}{e^{\omega\beta_{\rm g}}-1}).
\end{equation}
Plugging in Eq. \eqref{Expval-anticomm}, we find, for the thermal initial state,
\begin{equation*}
\begin{split}
    \expval{\acomm{q_\omega(t)}{q_\omega(t')}}_{\rm g}&=\frac{1}{m\omega}\cos\omega(t-t') \\
    &\hspace{0.2cm}+\frac{2}{m\omega}\frac{1}{e^{\omega\beta_{\rm g}}-1}\cos\omega(t-t'),
\end{split}
\end{equation*}
and the noise kernel becomes
\begin{equation} \label{Noise-kernel-thermal-state}
\begin{split}
    N_{\textrm{g (th)}}^{ijkl}(t,t')&=N_{\textrm{g (vac)}}^{ijkl}(t,t') \\
    &\hspace{0.2cm}+\frac{8m_0^2\pi^5}{\beta_{\rm g}^6}\mathcal{P}^{ijkl}F_{\rm th}\qty[\frac{\pi(t-t')}{\beta_{\rm g}}],
\end{split}
\end{equation}
where
\begin{equation} \label{Fth(x)}
\begin{split}
    F_{\rm th}(x)&\equiv\frac{1}{x^6} \\
    &\hspace{0.2cm}-\frac{1}{15\sinh^6x}\qty(2\cosh^4x+11\cosh^2x+2).
\end{split}
\end{equation}
\vspace{0.1cm}

\subsection{Coherent state}

For a quantum harmonic oscillator, the coherent state is defined by
\begin{equation}
    \ket{\alpha_\omega}=\mathscr{D}(\alpha_\omega)\ket{0},
\end{equation}
where $\mathscr{D}(\alpha_\omega)$ is called the \emph{coherent-state displacement operator}, defined as \cite{Loudon2000}
\begin{equation}
    \mathscr{D}(\alpha_\omega)=\exp(\alpha_\omega a_\omega^\dagger-\alpha_\omega^*a_\omega),
\end{equation}
with $a_{\omega}$ ($a_{\omega}^\dagger$) being the annihilation (creation) operator. Usually we take the displacement parameter to be independent of $\omega$, $\alpha_\omega=\alpha$. The displacement operator satisfies a number of properties,
\begin{equation}
\begin{split}
    \mathscr{D}^\dagger(\alpha)\mathscr{D}(\alpha)&=1, \\
    \mathscr{D}^\dagger(\alpha)a_\omega^n\mathscr{D}(\alpha)&=(a_\omega+\alpha)^n, \\
    \mathscr{D}^\dagger(\alpha)(a_\omega^\dagger)^n\mathscr{D}(\alpha)&=(a_\omega^\dagger+\alpha^*)^n,
\end{split}
\end{equation}
for $n\in\mathbb{N}$, from which we find
\begin{equation}
\begin{split}
    \expval{a_\omega^2}_{\rm g}&=\alpha^2, \\
    \expval{(a_\omega^\dagger)^2}_{\rm g}&=(\alpha^*)^2, \\
    \expval{H_\omega}_{\rm g}&=\omega\qty(\abs{\alpha}^2+\frac{1}{2}),
\end{split}
\end{equation}
where we also used the fact that the coherent state is an eigenstate of the annihilation operator, $a_{\omega}\ket{\alpha}=\alpha\ket{\alpha}$.

For simplicity, let us assume that the displacement parameter is real. In that case, we find
\begin{equation*}
\begin{split}
    \expval{\acomm{q_\omega(t)}{q_\omega(t')}}_{\rm g}&=\frac{1}{m\omega}\cos\omega(t-t') \\
    &\hspace{0.2cm}+\frac{\alpha^2}{m\omega}\cos(\omega t)\cos(\omega t'),
\end{split}
\end{equation*}
and the noise kernel becomes
\begin{equation} \label{Noise-kernel-coherent-state}
\begin{split}
    N_{\textrm{g (coh)}}^{ijkl}(t,t')&=N_{\textrm{g (vac)}}^{ijkl}(t,t') \\
    &+\frac{m_0^2\alpha^2}{15\pi}\mathcal{P}^{ijkl}\int_0^{\Tilde{\Lambda}_{\rm g}}d\omega\,\omega^5\cos(\omega t)\cos(\omega t'),
\end{split}
\end{equation}
where we regularized the divergent integral by introducing the momentum cutoff $\Tilde{\Lambda}_{\rm g}$.

\subsection{Squeezed state}

For a quantum harmonic oscillator, the squeezed state is defined by
\begin{equation}
    \ket{\zeta_\omega}=\mathscr{S}(\zeta_\omega)\ket{0},
\end{equation}
where $\mathscr{S}(\zeta_\omega)$ is called the \emph{squeeze operator}, defined as \cite{Loudon2000}
\begin{equation}
    \mathscr{S}(\zeta_\omega)=\exp[\frac{1}{2}\zeta_\omega^*a_\omega^2-\frac{1}{2}\zeta_\omega (a_\omega^\dagger)^2],
\end{equation}
where $\zeta_\omega$ is the \emph{complex squeeze parameter}
\begin{equation}
    \zeta_\omega=r_\omega e^{i\varphi_\omega}.
\end{equation}

Here we take the squeeze parameter to be independent of $\omega$, $\zeta_\omega=\zeta$. The squeeze operator satisfies a number of properties,
\begin{equation}
\begin{split}
    \mathscr{S}^\dagger(\zeta)\mathscr{S}(\zeta)&=1, \\
    \mathscr{S}^\dagger(\zeta)a_\omega^n\mathscr{S}(\zeta)&=(a\cosh r-a^\dagger e^{i\varphi}\sinh r)^n, \\
    \mathscr{S}^\dagger(\zeta)(a_\omega^\dagger)^n\mathscr{S}(\zeta)&=(a^\dagger\cosh r-a\,e^{-i\varphi}\sinh r)^n,
\end{split}
\end{equation}
for $n\in\mathbb{N}$, from which we find
\begin{equation}
\begin{split}
    \expval{a_\omega^2}_{\rm g}&=-e^{i\varphi}\sinh r\cosh r, \\
    \expval{(a_\omega^\dagger)^2}_{\rm g}&=-e^{-i\varphi}\sinh r\cosh r, \\
    \expval{H_\omega}_{\rm g}&=\frac{\omega}{2}\cosh2r,
\end{split}
\end{equation}
so that
\begin{equation*}
\begin{split}
    \expval{\acomm{q_\omega(t)}{q_\omega(t')}}_{\rm g}&=\frac{\cosh2r}{m\omega}\cos\omega(t-t') \\
    &\hspace{0.2cm}-\frac{\sinh2r}{m\omega}\cos\qty[\varphi-\omega(t+t')],
\end{split}
\end{equation*}
and the noise kernel becomes
\begin{equation} \label{Noise-kernel-squeezed-state}
\begin{split}
    N_{\textrm{g (sq)}}^{ijkl}(t,t')&=\cosh2r\,N_{\textrm{g (vac)}}^{ijkl}(t,t') \\
    &\hspace{0.2cm}-\frac{m_0^2\Bar{\Lambda}_{\rm g}^6}{15\pi}\sinh2r\,\mathcal{P}^{ijkl}F\qty[\Bar{\Lambda}_{\rm g}(t+t');\varphi],
\end{split}
\end{equation}
where we introduced the the momentum cutoff $\Bar{\Lambda}_{\rm g}$ and defined
\begin{equation}
\begin{split}
    &F(x;\varphi)\equiv\frac{1}{x^6}\left[ \qty(5x^4-60x^2+120)\cos(x-\varphi) \right. \\
    &\hspace{0.2cm}\left. +x\qty(x^4-20x^2+120)\sin\qty(x-\varphi)-120\cos\varphi\right] .
\end{split}
\end{equation}

%%%%%%%%%%%%%%%%%%%%%%%%%%%%%%%%%%%%%%%%%%%%%%%%
%%%%%%%%%%%%%%%%%%%%%%%%%%%%%%%%%%%%%%%%%%%%%%%%
%%%%%%%%%%%%%%%%%%%%%%%%%%%%%%%%%%%%%%%%%%%%%%%%
%%%%%%%%%%%%%%%%%%%%%%%%%%%%%%%%%%%%%%%%%%%%%%%%

\end{document}